\documentclass[draft]{agujournal2019}
\usepackage{url} 
\usepackage{lineno}
\usepackage[inline]{trackchanges} 
\usepackage{soul}
\usepackage[version=3]{mhchem}
\usepackage{amsmath}
\usepackage{setspace}
\journalname{arxiv.org}

\begin{document}

%
%


\title{A composite of the effects of major sudden stratospheric warming events on carbon dioxide radiative cooling in the mesosphere-lower-thermosphere}

%
%



\authors{Akash Kumar\affil{1}, M V Sunil Krishna\affil{1,2}, Alok K Ranjan\affil{1,3}}
	

\affiliation{1}{Department of Physics, Indian Institute of Technology Roorkee, Roorkee-247667, India}
\affiliation{2}{Centre for Space Science and Technology, Indian Institute of Technology Roorkee, Roorkee-247667, India}
\affiliation{3}{Space and Atmospheric Sciences Division, Physical Research Laboratory, Ahmedabad - 380009, India}
	




\correspondingauthor{M V Sunil Krishna}{mv.sunilkrishna@ph.iitr.ac.in}

\justify


\begin{keypoints}
\item A composite of major SSW events, and their impact on carbon dioxide (CO$_2$) radiative cooling is studied. 
\item CO$_2$ radiative cooling decreases and enhances during and after the Sudden Stratospheric Warming (SSW) events. 
\item Changes in oxygen density and temperature during SSW events primarily influence the CO$_2$ radiative cooling in the MLT region.
\end{keypoints}

%
%

%
%


\begin{abstract}

The major sudden stratospheric warming (SSW) events strongly influence the mean structure of the entire atmosphere, from the troposphere to the thermosphere. These events disrupt the compositional and thermal structure of the mesosphere and lower thermosphere (MLT), causing spatiotemporal variations in the concentration of trace species of this region. Currently, the role of dynamical changes during SSW events on radiative cooling in the MLT region is not well understood. An investigation of the SSW-induced changes in CO$_2$ radiative cooling in the MLT region is presented by examining the changes in the dynamics and transport of key species, such as CO$_2$ and atomic oxygen (O). A composite analysis has been performed to understand these changes during the major SSW events that occurred between 2005 and 2020. The variation of trace species is found to be associated with the change in vertical residual circulation. The results also show that CO$_2$ radiative cooling decreases during the mesospheric cooling that occurs during the stratospheric warming over the polar region. During the recovery stage of the SSW event, the CO$_2$ radiative cooling enhances in the mesosphere. These variations in CO$_2$ radiative cooling are mainly caused by temperature perturbations and oxygen transport in the MLT region. The contribution of temperature change and transport have also been investigated in detail.

\end{abstract}

\section*{Plain Language Summary}
Sudden Stratospheric Warming is a dynamic event that can cause large-scale changes in stratospheric temperature and polar stratospheric and mesospheric circulation pattern. These changes have been observed to be responsible for many physical chemical processes in the atmosphere starting from the troposphere up to the thermosphere. Carbon dioxide is an important trace species responsible for the radiative cooling of the MLT region. A composite analysis of changes in CO$_2$ radiative cooling during major SSW events that occurred \textcolor{black}{between 2005 and 2020 is presented} in this work. It has been found that the radiative cooling decreases during the SSW event over the polar region and its magnitude increases during the recovery of the SSW event. The SSW induced perturbations in temperature and oxygen density have been found to be responsible for the observed changes in radiative cooling intensities. 

\section{Introduction}

The Sudden Stratospheric Warming (SSW) events, considered as large-scale dynamical processes, are driven by the interaction of the stratospheric mean flow and upward-propagating planetary-scale waves from the lower atmosphere into the middle and upper atmosphere \cite{matsuno1971dynamical}. The overall probability of SSW occurrence depends on the climatology of both the amplitudes of the planetary-scale waves and the mean flow in the middle atmosphere. The winter polar stratosphere is dominated by a strong, westerly cold polar vortex. Wintertime westerlies in the stratosphere are sometimes slowed down or reversed by interactions between vertically propagating quasi-stationary planetary-scale waves and the stratospheric mean flow. This interaction causes wind anomaly that results in adiabatic heating in the polar stratosphere and interrupts the stratospheric polar vortex \cite{baldwin2021sudden, chandran2014stratosphere, andrews1987middle}.

The preliminary definition of an SSW event was given by the World Meteorological Organization (WMO, 1978), according to which a major SSW occurs when the latitudinal temperature increases poleward of 60$^\circ$ latitude and associated zonal circulation reverses at 60$^\circ$ latitude and 10 hPa pressure level. The warming events without the reversal of circulation at 10 hPa are referred to as minor SSW events. The major warming events that occurred in late winter are known as the final SSW events. Recent studies tried to expand the definition, scope, and conditions leading to the occurrence of the SSW events \cite{butler2015defining, charlton2007new, nath2016dynamics, kodera2016absorbing}. Most of the SSW events occur in the northern hemisphere, except two robust southern hemispheric warming events, which occurred in 2009 and 2019 \cite{baldwin2021sudden}. Although the SSW events are polar atmospheric phenomena occurring mainly in the northern hemisphere, their effects can extend to the southern hemisphere \cite{wang2019winter}. The decreasing temperature in the wintertime polar mesosphere associated with SSW events, known as mesospheric cooling, causes an immanent vertical instability in the region below the winter mesopause, which plays a crucial role in the exchange of mass and energy between the stratosphere and mesosphere \cite{shepherd2014stratospheric, limpasuvan2016composite}. The enhanced easterly gravity wave forcing following SSW events causes the polar stratopause to reform at mesospheric altitudes, known as Elevated stratopause.

 In the mesosphere and lower thermosphere (MLT), the meridional circulation is derived by integrated forcing of planetary waves (PWs), gravity waves (GWs), and their interaction with the mean flow \cite{laskar2019interhemispheric}. The reversal of winds from westerlies to easterlies takes place first in the mesosphere and upper stratosphere and then descends downward in the troposphere. The downward descent of temperature and wind anomalies impacts the lower atmosphere in the high- and mid-latitude troposphere during the subsequent weeks to months, influencing the surface weather \cite{butler2015defining}. Increasing evidences show that SSW events have substantial influences on the winds, temperature, and configuration of the atmosphere above and below the stratosphere due to the changes in zonal mean circulation and gravity wave filtering \cite{kurihara2010links, nayak2019variation, siskind2005observations, yiugit2016review}. During SSW events, the breakdown of the polar vortex enhances air mixing between the mid-latitudes and polar regions \cite{manney2009aura}. \textcolor{black}{Additionally, trace gases are often transported into a reformed and unusually strong vortex in the upper stratosphere and lower mesosphere following the event} \cite{bailey2014multi, damiani2010variability, de2018changes, kumar2024influence, liu2002study, randall2006enhanced, salmi2011mesosphere, shepherd2014stratospheric, zulicke2013structure}. During the northern hemispheric SSW events, \citeA{orsolini2022abrupt} also suggested changes in carbon dioxide (CO$_2$) \textcolor{black}{concentration} that were constrained to the higher latitudes.

Carbon dioxide plays an important role in controlling the thermal budget from the troposphere to the thermosphere. Its increasing concentration in the troposphere results in global warming leading to climate change \cite{lashof1990relative, solomon2009irreversible}. The increasing CO$_2$ concentration in the middle and upper atmosphere has been validated by recent studies \cite{garcia2016secular, lavstovivcka2023progress, qian2017carbon}. This increasing concentration can significantly impact the mean state of the middle atmosphere (30-110 km). Due to its prolonged lifespan (hundreds of days), CO$_2$ acts as a dynamical tracer \cite{chabrillat2002impact, wang2022climatology} and is also susceptible to alterations of the middle atmospheric circulation \cite{garcia2014distribution, lopez2000review}. Quantitatively, CO$_2$  has been identified as the most dominant radiative cooling agent in the middle atmosphere due to the infrared radiative emission by its vibrational relaxation \cite{kuhn1969infrared, castle2006vibrational, castle2012vibrational, mlynczak2010observations, mlynczak2022cooling, roble1989will}. The bands of CO$_2$ emission centered at 15 $\mu$m primarily dominate the radiative cooling in the MLT when compared to the other CO$_2$ emission bands \cite{houghton1969absorption, dickinson1984infrared}. The radiative cooling by CO$_2$ resulting in energy loss leads to the contraction of the upper atmosphere altering the atmospheric mean state \cite{akmaev2006impact, mlynczak2022cooling, roble1993greenhouse}. 

The radiative cooling is primarily caused by de-excitation of the CO$_2$ from its higher vibrational state to the lower vibrational state by means of spontaneous emission or by collision with other molecules, such as nitrogen and oxygen molecules (O$_2$ and N$_2$) and atomic oxygen (O) \cite{lopez2001non}. However, above $\sim80$ km, the role of O in the excitation and quenching of the vibrational mode of CO$_2$, and hence the radiative cooling, is extremely critical \cite{dickinson1984infrared, mlynczak2000contemporary, sharma1990role, shved2003measurement}. The atomic oxygen is also affected by these SSW events \cite{liu2002study, shepherd2010mesospheric}. Therefore, it is expected that variation in O can lead to changes in the CO$_2$-O collision excitation/quenching during these events. Therefore, variation in the trace species, such as CO$_2$ and O, in combination with temperature change in the MLT region, can significantly influence the cooling during the SSW events. \citeA{kumar2024effect} found an anti-correlation between CO$_2$ radiative cooling and mesospheric temperature while studying the influence of a major SSW event on CO$_2$ radiative cooling. They proposed that temperature variations, in addition to vertical transport of O, are responsible for variations in the population ratio of the CO$_2$ excited state, which in turn leads to the observed changes in CO$_2$ radiative cooling during SSW event. \textcolor{black}{While the responses of atomic oxygen and temperature during SSWs in the MLT region are relatively well studied, their combined effect on CO$_2$ radiative cooling remains inadequately understood. Therefore, this study aims to provide a robust characterization of CO$_2$ radiative cooling variations in response to changes in temperature, atomic oxygen, and CO$_2$ density during various major SSW events in the Northern Hemisphere.}

The present study focuses on investigating the changes in trace species like CO$_2$, O, and their impact on CO$_2$ infrared radiative emission rates for 15 $\mu$m band emission during several major SSW events, which occurred between 2005 and 2020. \textcolor{black}{The composite analysis approach considering various major SSW events aimed to develop a more comprehensive understanding of their atmospheric effects.} As stated earlier, CO$_2$ is the main contributor to mesospheric cooling, and its variability can significantly influence the thermal structure of the middle atmosphere. Therefore, the variability of CO$_2$ radiative emission rates and other important trace species responsible for the observed dynamical variations in resulting radiative cooling during the SSW events has been analyzed. These variabilities can help us understand the change in the energy budget of the MLT region and the coupling between different atmospheric layers.

\section{Data and Methodology}
 To identify the major SSW events that occurred between 2005 and 2020, the daily temperature and zonal mean wind data from Modern-Era Retrospective Analysis for Research and Applications, Version 2 (MERRA-2) reanalysis output have been analyzed. MERRA-2 is built with version 5.12.4 of NASA’s Global Modeling and Assimilation Office (GMAO) atmospheric data assimilation system. It has a horizontal resolution of approximately 0.5$^\circ$ × 0.625$^\circ$ in the latitude-longitude grid with 72 hybrid-eta levels from the surface to 0.01 hPa. The basic features of the MERRA-2 system are the same as the MERRA system \cite{rienecker2011merra}; however, numerous updates have been made to the previous version. A detailed description of MERRA-2 is presented in \citeA{gelaro2017modern}. In the present study, the zonal-mean temperature and zonal wind data are analyzed to identify the SSW occurrence. As suggested by \citeA{andrews1987middle}, a poleward temperature gradient and zonal-mean zonal wind reversal at 60$^\circ$ are used to mark the occurrence of the SSW, and eight major SSW events have been identified between 2005 and 2020. 
 
 To study the influence of the major SSW events on the dynamics, energetics and composition change in the MLT region, the observational data of temperature, CO$_2$ radiative cooling by 15 $\mu m$ band and O density derived from NASA's TIMED/SABER (Thermosphere, Ionosphere, Mesosphere, Energetics and Dynamics/Sounding of the Atmosphere using Broadband Emission Radiometry) has been used. \citeA{russell1999overview} and \citeA{esplin2023sounding} described the instrument’s features in detail. The temperature data, retrieved from CO$_2$ 15 $\mu$m emission, is available from $\sim$ 15 km to $\sim$ 110 km, O density is available from $\sim$ 80 km to $\sim$ 100 km, and CO$_2$ radiative cooling data available between 30-140 km in the SABER database. \textcolor{black}{The level 2A files of SABER with version 2.07 (V2.07) provide the vertical profiles of temperature and O density, and SABER V2.07 level 2B files contain CO$_2$ IR radiative cooling rates. The errors in SABER V2.07 temperature are 0.3–1.0 K at 30–70 km, 1.4 K at 80 km, 3.3 K at 90 km, 5.4 K at 100 km, and 10.5 K at 110 km.} The CO$_2$ radiative cooling rates used in the present analysis is selected mainly for the 15 $\mu m$ bands, particularly fundamental, first and second hot bands of the main isotope ($^{12}C^{16}O_2$) because it is the primary contributor to the CO$_2$ radiative cooling in the MLT region \cite{dickinson1984infrared}. In the present study, the above-mentioned datasets have been employed in the north viewing mode of SABER during the identified winters with major SSW events. 

 The meridional and vertical components of wind have been obtained from the specified-dynamics (SD) version of the thermospheric and ionospheric extension of the Whole Atmosphere Community Climate Model (WACCM-X) \cite{liu2018development}. With a peak pressure of $4.1\times10^{-10}$ hPa, the model has an identical vertical resolution below 0.96 hPa as the standard version of this model (WACCM), but it extends to 500–700 km altitude.  Above 0.96 hPa, vertical resolution has been increased to one-quarter scale height. \citeA{liu2018development} described the WACCM-X version 2.0 in detail. The daily averaged model output from SD WACCM-X simulation with nudging of MERRA-2 from the surface up to around 50 km is accessible through the Climate Data Gateway of the National Center for Atmospheric Research (NCAR). The vertical component of mean meridional residual circulation zonally averaged between 60$^{\circ}$ and 70$^{\circ}$ N, has been calculated using the scheme described in \citeA{andrews1987middle}. The SD WACCM-X output is used to create the composite of the vertical residual circulation  (w*) for the identified SSW events to investigate the vertical transport in the MLT.
 
 To understand the changes in the energetics of the MLT region during major SSW events, CO$_2$ concentration has been utilized from the observation of Atmospheric Chemistry Experiment Fourier Transform Spectrometer (ACE-FTS). ACE-FTS is composed of a high-resolution Fourier transform spectrometer and two imaging detectors onboard the SCISAT-1 satellite. A detailed description of the instruments is given in \citeA{soucy2002ace}, and the CO$_2$ mixing ratio retrieval process is documented in \citeA{foucher2011carbon}. \textcolor{black}{The method for retrieving CO$_2$ mixing ratio profiles from ACE-FTS observations involves selecting optimized spectral microwindows to minimize saturation and maximize accuracy. A two-step constrained least-squares approach and radiative transfer calculations are used, first retrieving temperature and pressure, then CO$_2$ concentrations, with regularization based on expected atmospheric variability. Using 15 microwindows reduces retrieval error to around 3 ppm, and vertical resolution is about 2–2.5 km. A correction is applied to align the lower-altitude isotopologue data with the main CO$_2$ mixing ratio profile, and final profiles are smoothed on a 1 km grid. Averaging over multiple occultations improves accuracy, with typical retrieval biases of 1 ppm and standard deviations of 2–4 ppm depending on altitude and season.} For the present study, CO$_2$ mixing ratio in parts per million (ppm) is derived from the ACE-FTS Level 2 Version 4.1/4.2 dataset in the northern polar region (60$^{\circ}$ - 70$^{\circ}$ N). The CO$_2$ density (cm$^{-3}$) is derived from the mixing ratio profile using the scheme described in \citeA{finlayson1999chemistry}. Zonal-mean daily variation of CO$_2$ density is used to study the impact of major SSW events. 

A composite analysis based on the identified major SSW events in the northern hemisphere from 2005 to 2020 is presented to construct a statistically robust image of the SSW characteristics and their impact on the MLT region. The duration of the major SSW events is separated into three phases, which are: (1) the pre-SSW phase, which spans for one week during which no significant changes in the stratospheric temperature and winds are observed; (2) the SSW occurrence/main phase, during which the stratosphere temperature rises, and zonal wind reverses; and (3) post-SSW phase, during which the stratospheric temperature and winds recover. Although a composite indicates the collective response of the selected events, there are some cases with minor SSW events preceding the major SSW events. Therefore, two cases have also been analyzed to identify the differences in the SSW responses between a sole major SSW event, occurred in 2009, and a major SSW event with a minor SSW event, occurred in 2010. \textcolor{black}{The 2009 SSW event was a strong major warming, with onset around 20 January and peak stratospheric temperatures occurring on 24 January 2009. This event significantly impacted the atmosphere across a broad range of altitudes. In contrast, during 2010, a minor SSW occurred around 20 January, followed by a major SSW event in early February, which also led to notable dynamical changes in the middle and upper atmosphere.}

\section{Results and discussion}

To examine the influence of the major SSW events on the CO$_2$ radiative cooling in the MLT, a composite analysis of eight major SSW events that happened in the northern hemisphere winters from 2005 to 2020, has been carried out. The duration choice for the current analysis is based on the availability of simultaneous observational data from the ACE-FTS, TIMED-SABER, and simulation output from SD WACCM-X. An SSW event is indicated by a poleward temperature gradient in the lower stratosphere; nevertheless, it is considered a major SSW event when the zonal mean zonal wind reversals at 60$^{\circ}$ latitude and 10 hPa level. 
Figure 1 shows the temperature difference between 60$^{\circ}$ N and 90$^{\circ}$ N and zonal-mean zonal wind at 60$^{\circ}$ N for all the winters with major SSW events from 2005 to 2020.
\begin{figure}
	\centering
	\noindent\includegraphics[scale=0.7]{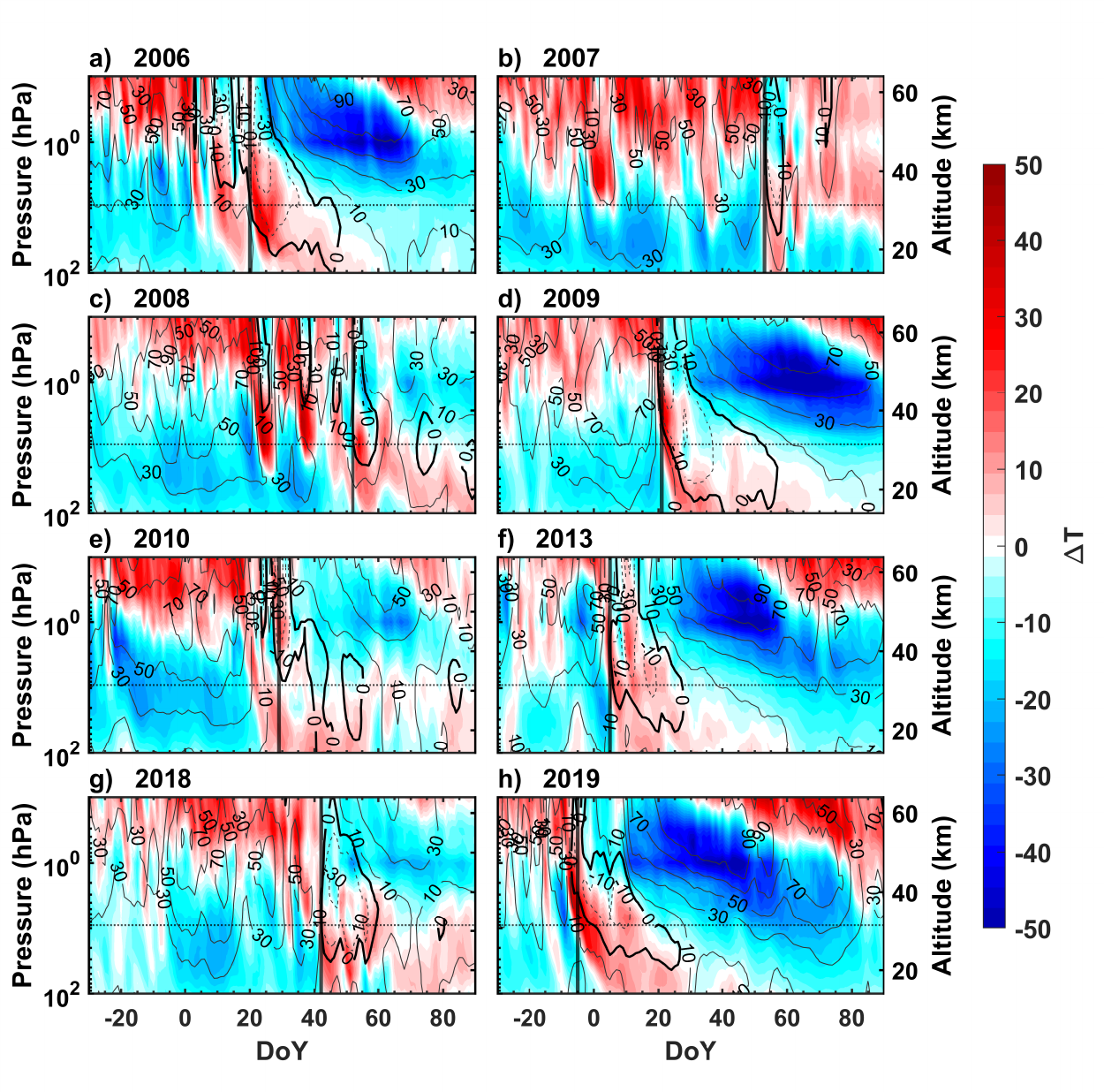}
	\caption{Zonal-mean daily averaged temperature difference (coloured map) over the northern polar region (between 90$^{\circ}$ and 60$^{\circ}$) and zonal-mean zonal wind (contour lines) at the defining latitude (60$^{\circ}$) derived from MERRA-2. The horizontal dashed line indicates the 10 hPa level, and the vertical line indicates the zonal-mean zonal wind reversal at 10 hPa level, indicating the SSW onset. \textcolor{black}{DoY indicate day of year with DoY 1 as 1 January of the corresponding year.}}
\end{figure}
 It is evident that there is a large and sudden positive temperature gradient over the northern polar region around the day of zonal-mean zonal wind reversal, therefore qualifying a major SSW event criterion. The vertical black line in Figure 1 indicates the central day (day 0) of the composite in our analysis when there was a strong poleward positive temperature gradient around the day of zonal-mean zonal wind reversal around 10 hPa level. 
 
The positive temperature gradient around 10 hPa level with no wind reversal before and after the central day represents the minor SSW events. There are some cases of major SSW events preceding the minor SSW events, e.g., 2006, 2007, 2008, 2010, and 2018, 2019, and some cases of sole SSW events, e.g., 2009 and 2013. As the poleward temperature gradient became positive in the lower stratosphere, there was also a negative poleward temperature gradient in the upper stratosphere and lower mesosphere after the major SSW event, as seen in the cases of 2006, 2009, 2013, and 2019 SSW events. Both the positive and negative poleward temperature gradient gradually descend into the lower stratosphere. It can also be noted that the positive temperature gradient gradually re-established in the upper stratosphere after the SSW events.

It is well established that the main cause of the positive temperature gradient and zonal-mean wind reversal is the breakdown of planetary waves having large vertical amplitudes. As suggested by \citeA{matsuno1971dynamical}, the planetary waves with large amplitude saturate in the higher latitude region constituting a critical level, and eventually induce the wind reversal there. The succeeding waves also saturate at this critical level that gradually propagate downward, and results in the sudden warming in the polar stratosphere. There is a simultaneous temperature decrease in the polar mesosphere due to changes in the gravity wave forcing and wind reversal \cite{limpasuvan2016composite}. These changes in the temperature and circulation can also affect the mean structure of the middle atmosphere, which needs to be studied for a comprehensive understanding of the vertical coupling of different atmospheric layers. 

To understand the influence of the major SSW events on the MLT region, a composite of temperature, the vertical component of the residual mean meridional circulation (w*, or simply vertical residual circulation), O, and CO$_2$ densities has been made and illustrated in Figure 2. 
\begin{figure}
	\centering
	\noindent\includegraphics[scale=0.6]{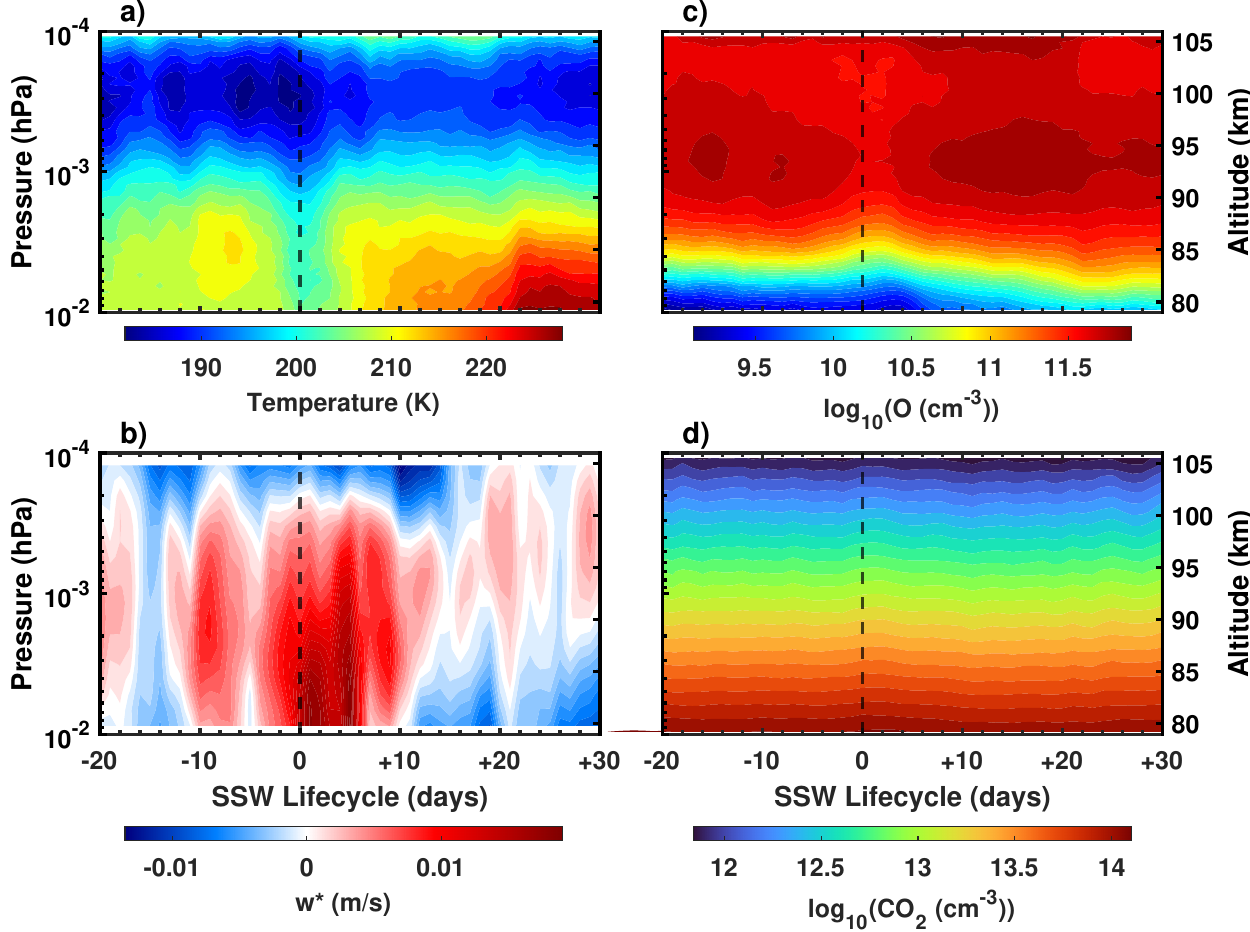}
	\caption{A composite of SABER-derived daily averaged temperature (a), SD WACCM-X derived vertical residual circulation (w*) (b), O density derived from SABER (c), and CO$_2$ density derived from ACE-FTS (d), zonally averaged between 60$^{\circ}$ - 70$^{\circ}$ N, during major SSW events that occurred between 2005 and 2020. The vertical dashed line indicates the central day of the composite.}
\end{figure}
It can be noted from the figure that there was a simultaneous and sharp decrease in temperature and O density (Figures 2a and 2c), a significant enhancement in w* (Figure 2b), and a minor increase in CO$_2$ density (Figure 2d) during the major SSW events, around the day 0. After the SSW event, the temperature and O were enhanced, after day +8, and there is a minor decrease in w* and CO$_2$ density in the recovery phase of SSW events. The maximum enhancement in the temperature is observed below 90 km, which is consistent with the downward residual circulation after the SSW events (after day +10). An increase in O density and a minor reduction in the CO$_2$ density is observed throughout the MLT after the SSW events.

These results are consistent with the previous studies suggesting the changes in the thermal and compositional structure of the MLT region \cite{tweedy2013nighttime, kumar2024effect}. The planetary wave-induced wave forcing resulted in changes in the mean meridional circulation in the middle atmosphere \cite{limpasuvan2016composite}. During the SSW events, the upward residual circulation in the MLT region causes upwelling and the expansion of the airmass, thereby adiabatic cooling in the mesosphere. The vertical residual circulation is very crucial for the distribution of minor constituents in the middle atmosphere \cite{de2018changes}. CO$_2$ rich and O poor airmass reaches higher altitudes in the polar region resulting in the CO$_2$ density enhancement and O density reduction during the SSW events. After the SSW events, the altered wave forcing in the MLT region resulted in enhanced downwelling, leading to temperature enhancement in the mesosphere and vertical transport of the trace species. The downwelling caused the CO$_2$ poor and O-rich airmass to reach the lower altitudes, leading to the observed depletion in CO$_2$ density and enhancement in O density after the SSW events. As mentioned earlier, these observed changes in the O and CO$_2$ densities, along with the temperature variations, can modify the CO$_2$ radiative cooling in the MLT region.

Figure 3 represents the composite of the daily variation in temperature, O density, CO$_2$ density, and CO$_2$ radiative cooling at 0.003 hPa ($\sim$ 85 km). 
\begin{figure}
	\centering
	\noindent\includegraphics[scale=0.8]{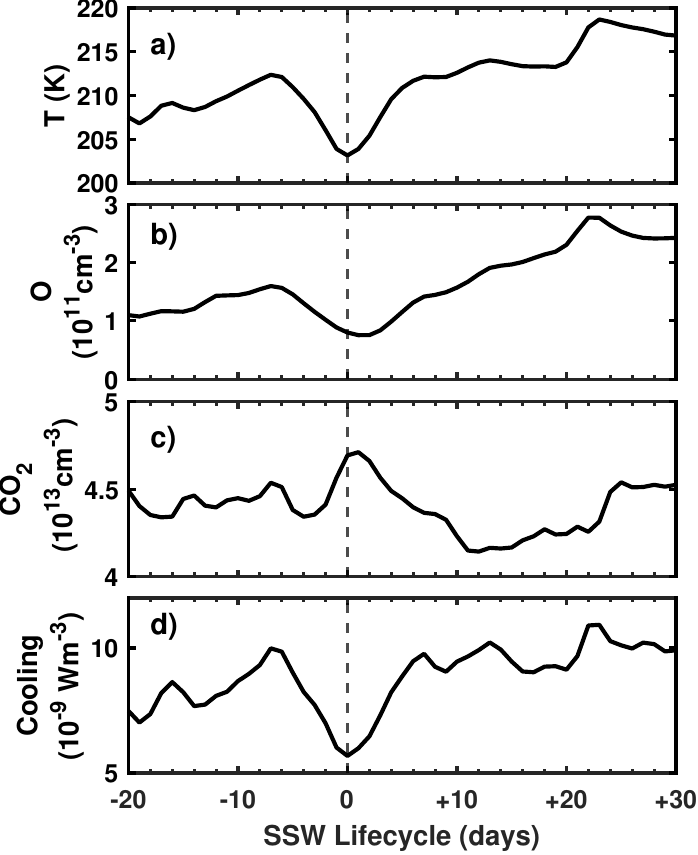}
	\caption{A composite of the SABER-derived temperature (a), O density (b), ACE-FTS derived CO$_2$ density (c), and \textcolor{black}{SABER-derived CO$_2$ radiative cooling (d),} between 60$^{\circ}$ - 70$^{\circ}$ N at 0.001 hPa. The vertical line indicates the central day of the composite.}
\end{figure}
It can be seen from Figure 3 that there was a simultaneous and sharp decrease in the temperature (Figure 3a), O density (Figure 3b), and CO$_2$ radiative cooling (Figure 3d),  and a significant enhancement in CO$_2$ density (Figure 3c) during the major SSW events. The patterns of decrease in temperature, O density, and CO$_2$ radiative cooling during the major SSW events are similar. The minor decrease prior to the major SSW events, around day -15, is due to the minor SSW events preceding these major SSW events, as mentioned earlier. The present analysis focuses on the effects only due to major SSW events. After the SSW events, the temperature, O density and CO$_2$ radiative cooling was gradually increased while CO$_2$ density was decreased till day +20. It can be noted that during the occurrence of the SSW, temperature, O density, and CO$_2$ radiative cooling were reduced by approximately 4\%, 50\%, and 35\%, respectively. In contrast, CO$_2$ density was enhanced by approximately 7\% compared to its mean values prior to the SSW event (from day -15 to day -8) at the 0.003 hPa level. After the occurrence of the SSW (from day +10 to day +25), a gradual enhancement of approximately 5\%, 90\%, and 25\% in temperature, O density, and CO$_2$ radiative cooling, respectively, as well as a 5\% reduction in CO$_2$ density, has been observed relative to their pre-SSW values. These parameters gradually begin to return to their pre-SSW values after day +25.

The reduced CO$_2$ radiative cooling during SSW occurrence can be attributed to reduced temperature and O density. Despite the minor enhancement in CO$_2$ density the large depletion in O density can be deduced to be the observed decrease in the CO$_2$ radiative cooling along with the mesospheric cooling. Due to its long lifetime, CO$_2$ can be influenced by dynamics, therefore its decreased/increased concentration in the MLT region indicate the upward/downward motion during/after the SSW events, as seen in Figure 2b. As mentioned earlier, the mesospheric cooling and the transport of trace species are mainly caused by the change in mean meridional circulation induced by change in the wave forcing. The decreased temperature and O density in the mesosphere results in lower collisional excitation of the CO$_2$ into its vibrational excited state, which explains the reduced CO$_2$ cooling during the initial phase of SSW event. The enhanced temperature in the polar mesosphere, caused by elevated stratopause re-appearing at the mesospheric altitudes in some of the major SSW events, and large O density are responsible for the higher collisional excitation and resulting in increased cooling after the SSW occurrence. It is evident from the discussion above that changes in circulation and wave dynamics result in variation in the key physical parameters and neutral number densities associated with the production mechanism of 15 $\mu$m band emission. It is important to establish the relative contribution of changes in the temperature, O and CO$_2$ density in the observed variability of CO$_2$ radiative cooling.
 
To further understand the variations in these atmospheric quantities and their relative importance, two major SSW events occurred during 2009 and 2010 are selected. The variation in vertical residual circulation (w*), temperature, O density, CO$_2$ density, and CO$_2$ radiative cooling, zonally averaged between 60$^\circ$ N and 70$^\circ$ N at 0.003 hPa level for 2009 (left column) and 2010 (right column) SSW events can be seen in figure 4.
\begin{figure}
	\centering
	\noindent\includegraphics[scale=0.6]{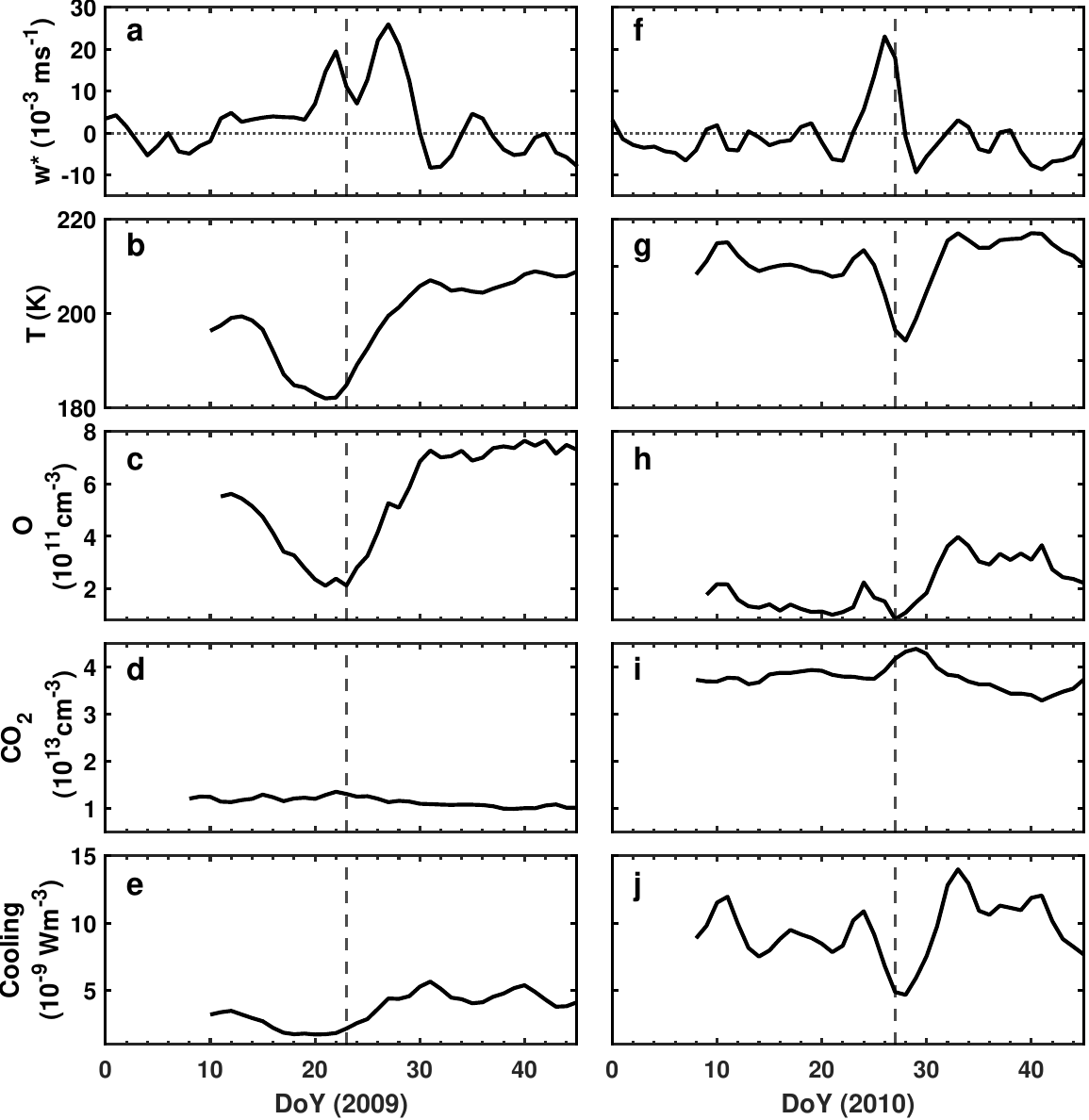}
	\caption{Daily variation in vertical residual circulation (a,f), temperature (b,g) and O density (c,h), CO$_2$ density (d,i), and CO$_2$ radiative cooling (e,j), zonally averaged between 60$^\circ$ N and 70$^\circ$ N during 2009 (left column) and 2010 (right column) major SSW events. The vertical line indicates the central day of strong positive temperature gradient in the lower stratosphere. \textcolor{black}{DoY indicate day of year with DoY 1 as 1 January of the corresponding year.}}
\end{figure}
A sudden increase in w*, indicating an upwelling, is noticed between DoY 20 and 30 in both years. During the 2009 SSW event, there was a sharp reduction of approximately 9\%, 64\%, and 50\% in temperature, O density, and CO$_2$ radiative cooling, respectively, as well as a 15\% increase in CO$_2$ density, compared to their mean values prior to the SSW event (between DoY 10 and 15). After the 2009 SSW event, temperature, O density, and CO$_2$ radiative cooling showed enhancements of approximately 5\%, 38\%, and 55\%, respectively, while CO$_2$ density increased by approximately 15\%, relative to their mean values before the SSW event. The quantities observed during both SSW events exhibit a similar variation as noted in the composite, although there are differences in the magnitude of these variations.
 
It is important to note that the 2010 major SSW event was preceded by a minor SSW event (arround DoY 20 in Figure 1e), the levels of CO$_2$ radiative cooling, temperature and O density were already depleted, and CO$_2$ density was slightly high before the major SSW occurrence. Therefore, one week period before the minor SSW occurrence has been considered as the pre-SSW period (between Doy 10 and 15). There was a significant decrease of 8\%, 40\%, and 45\% in temperature, O density, and CO$_2$ radiative cooling, respectively, and an enhancement of approximately 13\% in CO$_2$ density during the 2010 major SSW event, compared to their pre-SSW values, regardless of the minor SSW occurrence. After the major SSW event, an increase of approximately 3\%, 185\%, and 60\% in temperature, O density, and CO$_2$ radiative cooling, respectively, was observed, along with approximately 14\% reduction in CO$_2$ density in the polar MLT region, compared to their pre-SSW levels. The variations in these quantities were comparable, except for O density, during both SSW events.

The 2009 major SSW was a prolonged event causing elevated stratopause formation at mesospheric altitude and increased temperature there after the SSW event. The delayed downwelling after the extended upwelling also resulted in transport of O density and decreased CO$_2$ density into the mesosphere, after the SSW event. In the case of 2010 major SSW event, the upwelling was limited in time, which could be due to prior episode of minor SSW event that already disturbed the mean structure of the middle atmosphere associated with minor upwelling and downwelling in the mesosphere. A minor yet sharp enhancement in CO$_2$ radiative cooling, temperature and O density, and a small depletion in CO$_2$ density is due to the enhanced downwelling around DoY 40 after 2010 major SSW event. The variations in polar mesospheric temperature and vertically transported O density are expected to cause the sudden changes in the CO$_2$ radiative cooling during and after the SSW events. Therefore, it is important to estimate the relative contribution of changes in the temperature, O and CO$_2$ densities during and after the major SSW events has been investigated to know which parameter is dominating the variation in CO$_2$ radiative cooling during the major SSW event.
 
The variations in CO$_2$ radiative cooling can also be estimated from the emission rates for the 15 $\mu$m band, as described in \cite{khomich2008airglow}, because it represents the ratio of the rates of excitation and de-excitation of CO$_2$ to its vibrational excited state. The emission rate variability at 0.0035 hPa, and contribution of various parameters, has been analyzed for the composite of the SSW events, and presented in figure 5.
\begin{figure}
	\centering
	\noindent\includegraphics[scale=1]{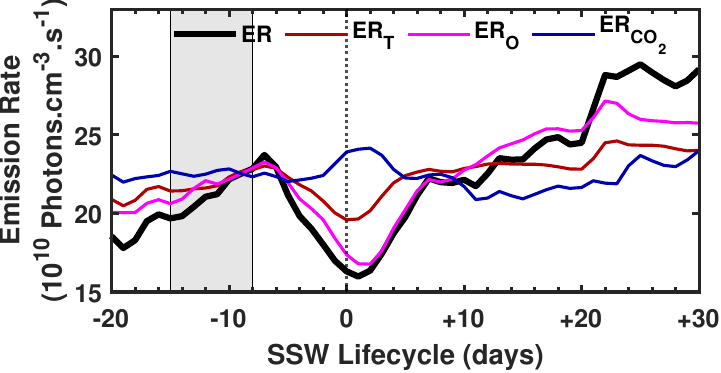}
	\caption{A composite of the daily variation in emission rates of 15 $\mu$m band from CO$_2$ (black line), as a function of changes in temperature only (red line), O density only (magenta line) and CO$_2$ density only (blue line), zonally averaged between 60$^\circ$ N and 70$^\circ$ N. The shaded region indicate the pre-SSW period and vertical line indicates the central day of composite.}
\end{figure}
The CO$_2$ 15 $\mu$m emission rate variations and relative contributions due to a specific parameter (e.g., temperature, O or CO$_2$ densities) have been analyzed by keeping other parameters constant at their pre-SSW values (shaded region). For instance, in the emission rates, considering the changes in temperature only and keeping O and CO$_2$ densities constant to their pre-SSW values provides the contribution due to temperature. Similarly, the changes in emission rates with respect to O density are analyzed while keeping the temperature constant to its pre-SSW values. It can be seen from the figure that the emission rate also exhibits similar variation as seen in the CO$_2$ radiative cooling during and after the SSW events. The overall emission rate decreases by $\sim25\%$ during the SSW occurrences and increases $\sim40\%$ after the SSW events in comparison to its pre-SSW mean values. The emission rate due to the only temperature variability and due to change in only O density also decreased by $\sim12\%$ and $\sim23\%$, respectively, while $\sim8\%$ increase in emission rate is seen due to change in only CO$_2$ density during the SSW events. After the SSW events, the emission rate due to only temperature change and due to only O density change increased by $\sim15\%$ and $\sim28\%$, respectively, and emission rate due to only \textcolor{black}{CO$_2$ density} changes decreases by $\sim10\%$ in comparison to their pre-SSW values. It can be noted that the magnitude of emission rate variability due to change in only O density is greater than due to change in only temperature, and the role of \textcolor{black}{CO$_2$ density} changes appeared to be limitted. The large variation in emission rates due to only O density changes in comparison to changes due to only temperature variability, indicate the important role of O density in controlling the 15 $\mu$m infrared emissions by CO$_2$.

The emission rates coincide with the population ratio (CO$_2$($01^10$)/CO$_2$($00^00$)), which depends on temperature and O density participating in collisional excitation/de-excitation of CO$_2$ in the MLT region \cite{castle2012vibrational}. As a result, it can be noted that despite minor variation in CO$_2$ density a depletion in O density and temperature leads to the observed changes in the emission rates and hence CO$_2$ radiative cooling in the MLT region during and after the SSW period. The simultaneous decrease in O density and the mesospheric temperature during the SSW occurrence result in the decreased collisional excitation of CO$_2$ into the higher vibrational excited state, and leading to reduced emission rates. As stated earlier, the downwelling after the SSW event also results in adiabatic heating and hence large mesospheric temperature, which in combination with the vertically transported O density result in enhanced collisional excitation of the CO$_2$ in its higher vibrational excited states and hence increased emission rates, which eventually results in enhanced CO$_2$ radiative cooling in the MLT region after the SSW event. 

To begin with, the role of temperature variation was expected to be the dominating in the CO$_2$ radiative cooling. \textcolor{black}{ It is expected from the chemistry that a small change in the temperature lead to large variations in emission rates because temperature has stronger influence on radiative cooling due to the exponential dependence. It has been found that large variations in O density along with the small temperature variation control the CO$_2$ radiative cooling in the MLT region during and after the SSW event}. Therefore, it is evident from the above discussion that a minor variation in CO$_2$ density during and after the SSW event does not have significant contribution in the observed CO$_2$ radiative cooling variability. The availability of atomic oxygen is crucial for the CO$_2$ radiative cooling along with the temperature variation. Therefore, It can be suggest that the observed variation in CO$_2$ radiative cooling is dynamically controlled primarily by atomic oxygen and temperature changes, which are caused by wave induced circulation changes during and after the SSW event.

\section{Summary}
This study investigates the impact of the major SSW events that occurred between 2005 and 2020 on the MLT region. A composite analysis has been presented to explore the observed changes in the CO$_2$ radiative cooling and associated parameters during the major SSW events in the considered time period. The wave-induced circulation change results in variations in the temperature and other trace species in the middle atmosphere during these events. The upwelling/downwelling induced by the variable vertical residual circulation results in the reduced/enhanced mesospheric temperature and O density in the polar MLT region during/after the SSW occurrence. The changes in vertical residual circulation also results in increased/decreased CO$_2$ density due to upwelling/downwelling in the polar MLT region during/after the SSW events. An opposite tendency between CO$_2$ density and CO$_2$ radiative cooling has been observed during and after the SSW event. The CO$_2$ radiative cooling was decreased during the SSW events and it was enhanced after the SSW event. This cooling pattern is consistent with the variation in temperature and O density, and opposite to changes in CO$_2$ density during the studied periods. The mesospheric temperature and O variability induced changes in the collisional excitation/de-excitation of the CO$_2$ to the vibrationally excited state, leads to a possible explanation for the observed changes. Therefore, it can be concluded that a large variability of oxygen density in the MLT region by vertical transport, induced by upwelling/downwelling, is very crucial along with temperature changes in the observed CO$_2$ radiative cooling variations, during the major SSW events. A comprehensive analysis of the roles of temperature and oxygen density in the radiative transfer during each SSW event is essential to strengthening our understanding of the energetic changes in the MLT region during these large-scale dynamic phenomena.

\section*{Data Availability Statement}
\textcolor{black}{The MERRA-2 data is obtained from \url{https://disc.gsfc.nasa.gov/datasets}} \cite{MERRA2_2015}. The TIMED/SABER data can be downloaded from the following link: \url{http://saber.gats-inc.com/data.php} \cite{SABER2023}. The WACCM-X output data, as free-run history files, can be obtained from the \textcolor{black}{Research Data Archive at NCAR at \url{https://rda.ucar.edu/datasets/d651034/} \cite{WACCM_X2023}}, and ACE-FTS data has been obtained from \url{http://www.ace.uwaterloo.ca/data.php} \cite{ACE_FTS}.

\justify

\acknowledgments
The authors acknowledge the MERRA2 team for making the data publicly available. The authors also thank the TIMED-SABER instrument, algorithm, and data processing teams for providing free access to the SABER dataset and the SCISAT/ACE-FTS teams for providing the data used in this study. We sincerely thank the WACCM team for providing free-run modelled data for WACCM-X. The authors thank the DST-SERB and the Ministry of Education, Government of India, for the research assistantship.

%

\bibliography{References}

\begin{thebibliography}{}

\bibitem [\protect \citeauthoryear {%
ACE-FTS-Team%
}{%
ACE-FTS-Team%
}{%
{\protect \APACyear {2003}}%
}]{%
ACE_FTS}
\APACinsertmetastar {%
ACE_FTS}%
\begin{APACrefauthors}%
ACE-FTS-Team.%
\end{APACrefauthors}%
\unskip\
\newblock
\APACrefYearMonthDay{2003}{}{}.
\newblock
\APACrefbtitle {{Atmospheric Chemistry Experiment Fourier Transform Spectrometer Dataset}.} {{Atmospheric Chemistry Experiment Fourier Transform Spectrometer Dataset}.}
\newblock
\APACaddressPublisher{}{Canadian Space Agency [Dataset]}.
\newblock
\begin{APACrefURL} \url{http://www.ace.uwaterloo.ca/data.php} \end{APACrefURL}
\PrintBackRefs{\CurrentBib}

\bibitem [\protect \citeauthoryear {%
Akmaev%
, Fomichev%
\BCBL {}\ \BBA {} Zhu%
}{%
Akmaev%
\ \protect \BOthers {.}}{%
{\protect \APACyear {2006}}%
}]{%
akmaev2006impact}
\APACinsertmetastar {%
akmaev2006impact}%
\begin{APACrefauthors}%
Akmaev, R.%
, Fomichev, V.%
\BCBL {}\ \BBA {} Zhu, X.%
\end{APACrefauthors}%
\unskip\
\newblock
\APACrefYearMonthDay{2006}{}{}.
\newblock
{\BBOQ}\APACrefatitle {Impact of middle-atmospheric composition changes on greenhouse cooling in the upper atmosphere} {Impact of middle-atmospheric composition changes on greenhouse cooling in the upper atmosphere}.{\BBCQ}
\newblock
\APACjournalVolNumPages{Journal of atmospheric and solar-terrestrial physics}{68}{17}{1879--1889}.
\newblock
\begin{APACrefDOI} \doi{https://doi.org/10.1016/j.jastp.2006.03.008} \end{APACrefDOI}
\PrintBackRefs{\CurrentBib}

\bibitem [\protect \citeauthoryear {%
Andrews%
, Holton%
\BCBL {}\ \BBA {} Leovy%
}{%
Andrews%
\ \protect \BOthers {.}}{%
{\protect \APACyear {1987}}%
}]{%
andrews1987middle}
\APACinsertmetastar {%
andrews1987middle}%
\begin{APACrefauthors}%
Andrews, D\BPBI G.%
, Holton, J\BPBI R.%
\BCBL {}\ \BBA {} Leovy, C\BPBI B.%
\end{APACrefauthors}%
\unskip\
\newblock
\APACrefYear{1987}.
\newblock
\APACrefbtitle {Middle atmosphere dynamics} {Middle atmosphere dynamics}\ (\BNUM~40).
\newblock
\APACaddressPublisher{}{Academic press}.
\PrintBackRefs{\CurrentBib}

\bibitem [\protect \citeauthoryear {%
Bailey%
\ \protect \BOthers {.}}{%
Bailey%
\ \protect \BOthers {.}}{%
{\protect \APACyear {2014}}%
}]{%
bailey2014multi}
\APACinsertmetastar {%
bailey2014multi}%
\begin{APACrefauthors}%
Bailey, S.%
, Thurairajah, B.%
, Randall, C.%
, Holt, L.%
, Siskind, D.%
, Harvey, V.%
\BDBL {}Russell, J.%
\end{APACrefauthors}%
\unskip\
\newblock
\APACrefYearMonthDay{2014}{}{}.
\newblock
{\BBOQ}\APACrefatitle {A multi tracer analysis of thermosphere to stratosphere descent triggered by the 2013 Stratospheric Sudden Warming} {A multi tracer analysis of thermosphere to stratosphere descent triggered by the 2013 stratospheric sudden warming}.{\BBCQ}
\newblock
\APACjournalVolNumPages{Geophysical Research Letters}{41}{14}{5216--5222}.
\newblock
\begin{APACrefDOI} \doi{https://doi.org/10.1002/2014GL059860} \end{APACrefDOI}
\PrintBackRefs{\CurrentBib}

\bibitem [\protect \citeauthoryear {%
Baldwin%
\ \protect \BOthers {.}}{%
Baldwin%
\ \protect \BOthers {.}}{%
{\protect \APACyear {2021}}%
}]{%
baldwin2021sudden}
\APACinsertmetastar {%
baldwin2021sudden}%
\begin{APACrefauthors}%
Baldwin, M\BPBI P.%
, Ayarzag{\"u}ena, B.%
, Birner, T.%
, Butchart, N.%
, Butler, A\BPBI H.%
, Charlton-Perez, A\BPBI J.%
\BDBL {}others%
\end{APACrefauthors}%
\unskip\
\newblock
\APACrefYearMonthDay{2021}{}{}.
\newblock
{\BBOQ}\APACrefatitle {Sudden stratospheric warmings} {Sudden stratospheric warmings}.{\BBCQ}
\newblock
\APACjournalVolNumPages{Reviews of Geophysics}{59}{1}{e2020RG000708}.
\newblock
\begin{APACrefDOI} \doi{https://doi.org/10.1029/2020RG000708} \end{APACrefDOI}
\PrintBackRefs{\CurrentBib}

\bibitem [\protect \citeauthoryear {%
Butler%
\ \protect \BOthers {.}}{%
Butler%
\ \protect \BOthers {.}}{%
{\protect \APACyear {2015}}%
}]{%
butler2015defining}
\APACinsertmetastar {%
butler2015defining}%
\begin{APACrefauthors}%
Butler, A\BPBI H.%
, Seidel, D\BPBI J.%
, Hardiman, S\BPBI C.%
, Butchart, N.%
, Birner, T.%
\BCBL {}\ \BBA {} Match, A.%
\end{APACrefauthors}%
\unskip\
\newblock
\APACrefYearMonthDay{2015}{}{}.
\newblock
{\BBOQ}\APACrefatitle {Defining sudden stratospheric warmings} {Defining sudden stratospheric warmings}.{\BBCQ}
\newblock
\APACjournalVolNumPages{Bulletin of the American Meteorological Society}{96}{11}{1913--1928}.
\newblock
\begin{APACrefDOI} \doi{https://doi.org/10.1175/BAMS-D-13-00173.1} \end{APACrefDOI}
\PrintBackRefs{\CurrentBib}

\bibitem [\protect \citeauthoryear {%
Castle%
, Black%
, Simione%
\BCBL {}\ \BBA {} Dodd%
}{%
Castle%
\ \protect \BOthers {.}}{%
{\protect \APACyear {2012}}%
}]{%
castle2012vibrational}
\APACinsertmetastar {%
castle2012vibrational}%
\begin{APACrefauthors}%
Castle, K\BPBI J.%
, Black, L\BPBI A.%
, Simione, M\BPBI W.%
\BCBL {}\ \BBA {} Dodd, J\BPBI A.%
\end{APACrefauthors}%
\unskip\
\newblock
\APACrefYearMonthDay{2012}{}{}.
\newblock
{\BBOQ}\APACrefatitle {Vibrational Relaxation of CO2 ($\nu$2) by O (3P) in the 142--490 K Temperature Range} {Vibrational relaxation of co2 ($\nu$2) by o (3p) in the 142--490 k temperature range}.{\BBCQ}
\newblock
\APACjournalVolNumPages{Journal of Geophysical Research: Space Physics}{117}{A4}{}.
\newblock
\begin{APACrefDOI} \doi{https://doi.org/10.1029/2012JA017519} \end{APACrefDOI}
\PrintBackRefs{\CurrentBib}

\bibitem [\protect \citeauthoryear {%
Castle%
, Kleissas%
, Rhinehart%
, Hwang%
\BCBL {}\ \BBA {} Dodd%
}{%
Castle%
\ \protect \BOthers {.}}{%
{\protect \APACyear {2006}}%
}]{%
castle2006vibrational}
\APACinsertmetastar {%
castle2006vibrational}%
\begin{APACrefauthors}%
Castle, K\BPBI J.%
, Kleissas, K\BPBI M.%
, Rhinehart, J\BPBI M.%
, Hwang, E\BPBI S.%
\BCBL {}\ \BBA {} Dodd, J\BPBI A.%
\end{APACrefauthors}%
\unskip\
\newblock
\APACrefYearMonthDay{2006}{}{}.
\newblock
{\BBOQ}\APACrefatitle {Vibrational relaxation of CO2 ($\nu$2) by atomic oxygen} {Vibrational relaxation of co2 ($\nu$2) by atomic oxygen}.{\BBCQ}
\newblock
\APACjournalVolNumPages{Journal of Geophysical Research: Space Physics}{111}{A9}{}.
\newblock
\begin{APACrefDOI} \doi{https://doi.org/10.1029/2006JA011736} \end{APACrefDOI}
\PrintBackRefs{\CurrentBib}

\bibitem [\protect \citeauthoryear {%
Chabrillat%
, Kockarts%
, Fonteyn%
\BCBL {}\ \BBA {} Brasseur%
}{%
Chabrillat%
\ \protect \BOthers {.}}{%
{\protect \APACyear {2002}}%
}]{%
chabrillat2002impact}
\APACinsertmetastar {%
chabrillat2002impact}%
\begin{APACrefauthors}%
Chabrillat, S.%
, Kockarts, G.%
, Fonteyn, D.%
\BCBL {}\ \BBA {} Brasseur, G.%
\end{APACrefauthors}%
\unskip\
\newblock
\APACrefYearMonthDay{2002}{}{}.
\newblock
{\BBOQ}\APACrefatitle {Impact of molecular diffusion on the CO2 distribution and the temperature in the mesosphere} {Impact of molecular diffusion on the co2 distribution and the temperature in the mesosphere}.{\BBCQ}
\newblock
\APACjournalVolNumPages{Geophysical research letters}{29}{15}{19--1}.
\newblock
\begin{APACrefDOI} \doi{https://doi.org/10.1029/2002GL015309} \end{APACrefDOI}
\PrintBackRefs{\CurrentBib}

\bibitem [\protect \citeauthoryear {%
Chandran%
, Collins%
\BCBL {}\ \BBA {} Harvey%
}{%
Chandran%
\ \protect \BOthers {.}}{%
{\protect \APACyear {2014}}%
}]{%
chandran2014stratosphere}
\APACinsertmetastar {%
chandran2014stratosphere}%
\begin{APACrefauthors}%
Chandran, A.%
, Collins, R.%
\BCBL {}\ \BBA {} Harvey, V.%
\end{APACrefauthors}%
\unskip\
\newblock
\APACrefYearMonthDay{2014}{}{}.
\newblock
{\BBOQ}\APACrefatitle {Stratosphere-mesosphere coupling during stratospheric sudden warming events} {Stratosphere-mesosphere coupling during stratospheric sudden warming events}.{\BBCQ}
\newblock
\APACjournalVolNumPages{Advances in Space Research}{53}{9}{1265--1289}.
\newblock
\begin{APACrefDOI} \doi{https://doi.org/10.1016/j.asr.2014.02.005} \end{APACrefDOI}
\PrintBackRefs{\CurrentBib}

\bibitem [\protect \citeauthoryear {%
Charlton%
\ \BBA {} Polvani%
}{%
Charlton%
\ \BBA {} Polvani%
}{%
{\protect \APACyear {2007}}%
}]{%
charlton2007new}
\APACinsertmetastar {%
charlton2007new}%
\begin{APACrefauthors}%
Charlton, A\BPBI J.%
\BCBT {}\ \BBA {} Polvani, L\BPBI M.%
\end{APACrefauthors}%
\unskip\
\newblock
\APACrefYearMonthDay{2007}{}{}.
\newblock
{\BBOQ}\APACrefatitle {A new look at stratospheric sudden warmings. Part I: Climatology and modeling benchmarks} {A new look at stratospheric sudden warmings. part i: Climatology and modeling benchmarks}.{\BBCQ}
\newblock
\APACjournalVolNumPages{Journal of climate}{20}{3}{449--469}.
\newblock
\begin{APACrefDOI} \doi{https://doi.org/10.1175/JCLI3996.1} \end{APACrefDOI}
\PrintBackRefs{\CurrentBib}

\bibitem [\protect \citeauthoryear {%
Damiani%
, Storini%
, Santee%
\BCBL {}\ \BBA {} Wang%
}{%
Damiani%
\ \protect \BOthers {.}}{%
{\protect \APACyear {2010}}%
}]{%
damiani2010variability}
\APACinsertmetastar {%
damiani2010variability}%
\begin{APACrefauthors}%
Damiani, A.%
, Storini, M.%
, Santee, M.%
\BCBL {}\ \BBA {} Wang, S.%
\end{APACrefauthors}%
\unskip\
\newblock
\APACrefYearMonthDay{2010}{}{}.
\newblock
{\BBOQ}\APACrefatitle {Variability of the nighttime OH layer and mesospheric ozone at high latitudes during northern winter: influence of meteorology} {Variability of the nighttime oh layer and mesospheric ozone at high latitudes during northern winter: influence of meteorology}.{\BBCQ}
\newblock
\APACjournalVolNumPages{Atmospheric Chemistry and Physics}{10}{21}{10291--10303}.
\newblock
\begin{APACrefDOI} \doi{https://doi.org/10.5194/acp-10-10291-2010} \end{APACrefDOI}
\PrintBackRefs{\CurrentBib}

\bibitem [\protect \citeauthoryear {%
de~la C{\'a}mara%
, Abalos%
\BCBL {}\ \BBA {} Hitchcock%
}{%
de~la C{\'a}mara%
\ \protect \BOthers {.}}{%
{\protect \APACyear {2018}}%
}]{%
de2018changes}
\APACinsertmetastar {%
de2018changes}%
\begin{APACrefauthors}%
de~la C{\'a}mara, A.%
, Abalos, M.%
\BCBL {}\ \BBA {} Hitchcock, P.%
\end{APACrefauthors}%
\unskip\
\newblock
\APACrefYearMonthDay{2018}{}{}.
\newblock
{\BBOQ}\APACrefatitle {Changes in stratospheric transport and mixing during sudden stratospheric warmings} {Changes in stratospheric transport and mixing during sudden stratospheric warmings}.{\BBCQ}
\newblock
\APACjournalVolNumPages{Journal of Geophysical Research: Atmospheres}{123}{7}{3356--3373}.
\newblock
\begin{APACrefDOI} \doi{https://doi.org/10.1002/2017JD028007} \end{APACrefDOI}
\PrintBackRefs{\CurrentBib}

\bibitem [\protect \citeauthoryear {%
Dickinson%
}{%
Dickinson%
}{%
{\protect \APACyear {1984}}%
}]{%
dickinson1984infrared}
\APACinsertmetastar {%
dickinson1984infrared}%
\begin{APACrefauthors}%
Dickinson, R\BPBI E.%
\end{APACrefauthors}%
\unskip\
\newblock
\APACrefYearMonthDay{1984}{}{}.
\newblock
{\BBOQ}\APACrefatitle {Infrared radiative cooling in the mesosphere and lower thermosphere} {Infrared radiative cooling in the mesosphere and lower thermosphere}.{\BBCQ}
\newblock
\APACjournalVolNumPages{Journal of atmospheric and terrestrial physics}{46}{11}{995--1008}.
\newblock
\begin{APACrefDOI} \doi{https://doi.org/10.1016/0021-9169(84)90006-0} \end{APACrefDOI}
\PrintBackRefs{\CurrentBib}

\bibitem [\protect \citeauthoryear {%
Esplin%
, Mlynczak%
, Russell%
, Gordley%
\BCBL {}\ \BBA {} Team%
}{%
Esplin%
\ \protect \BOthers {.}}{%
{\protect \APACyear {2023}}%
}]{%
esplin2023sounding}
\APACinsertmetastar {%
esplin2023sounding}%
\begin{APACrefauthors}%
Esplin, R.%
, Mlynczak, M\BPBI G.%
, Russell, J.%
, Gordley, L.%
\BCBL {}\ \BBA {} Team, S.%
\end{APACrefauthors}%
\unskip\
\newblock
\APACrefYearMonthDay{2023}{}{}.
\newblock
{\BBOQ}\APACrefatitle {Sounding of the Atmosphere using Broadband Emission Radiometry (SABER): Instrument and science measurement description} {Sounding of the atmosphere using broadband emission radiometry (saber): Instrument and science measurement description}.{\BBCQ}
\newblock
\APACjournalVolNumPages{Earth and Space Science}{10}{9}{e2023EA002999}.
\newblock
\begin{APACrefDOI} \doi{https://doi.org/10.1029/2023EA002999} \end{APACrefDOI}
\PrintBackRefs{\CurrentBib}

\bibitem [\protect \citeauthoryear {%
Finlayson-Pitts%
\ \BBA {} Pitts~Jr%
}{%
Finlayson-Pitts%
\ \BBA {} Pitts~Jr%
}{%
{\protect \APACyear {1999}}%
}]{%
finlayson1999chemistry}
\APACinsertmetastar {%
finlayson1999chemistry}%
\begin{APACrefauthors}%
Finlayson-Pitts, B\BPBI J.%
\BCBT {}\ \BBA {} Pitts~Jr, J\BPBI N.%
\end{APACrefauthors}%
\unskip\
\newblock
\APACrefYear{1999}.
\newblock
\APACrefbtitle {Chemistry of the upper and lower atmosphere: theory, experiments, and applications} {Chemistry of the upper and lower atmosphere: theory, experiments, and applications}.
\newblock
\APACaddressPublisher{}{Elsevier}.
\newblock
\begin{APACrefDOI} \doi{https://doi.org/10.1016/B978-0-12-257060-5.X5000-X} \end{APACrefDOI}
\PrintBackRefs{\CurrentBib}

\bibitem [\protect \citeauthoryear {%
Foucher%
\ \protect \BOthers {.}}{%
Foucher%
\ \protect \BOthers {.}}{%
{\protect \APACyear {2011}}%
}]{%
foucher2011carbon}
\APACinsertmetastar {%
foucher2011carbon}%
\begin{APACrefauthors}%
Foucher, P.%
, Ch{\'e}din, A.%
, Armante, R.%
, Boone, C.%
, Crevoisier, C.%
\BCBL {}\ \BBA {} Bernath, P.%
\end{APACrefauthors}%
\unskip\
\newblock
\APACrefYearMonthDay{2011}{}{}.
\newblock
{\BBOQ}\APACrefatitle {Carbon dioxide atmospheric vertical profiles retrieved from space observation using ACE-FTS solar occultation instrument} {Carbon dioxide atmospheric vertical profiles retrieved from space observation using ace-fts solar occultation instrument}.{\BBCQ}
\newblock
\APACjournalVolNumPages{Atmospheric Chemistry and Physics}{11}{6}{2455--2470}.
\newblock
\begin{APACrefDOI} \doi{https://doi.org/10.5194/acp-11-2455-2011} \end{APACrefDOI}
\PrintBackRefs{\CurrentBib}

\bibitem [\protect \citeauthoryear {%
Garcia%
\ \protect \BOthers {.}}{%
Garcia%
\ \protect \BOthers {.}}{%
{\protect \APACyear {2016}}%
}]{%
garcia2016secular}
\APACinsertmetastar {%
garcia2016secular}%
\begin{APACrefauthors}%
Garcia, R\BPBI R.%
, L{\'o}pez-Puertas, M.%
, Funke, B.%
, Kinnison, D\BPBI E.%
, Marsh, D\BPBI R.%
\BCBL {}\ \BBA {} Qian, L.%
\end{APACrefauthors}%
\unskip\
\newblock
\APACrefYearMonthDay{2016}{}{}.
\newblock
{\BBOQ}\APACrefatitle {On the secular trend of COx and CO2 in the lower thermosphere} {On the secular trend of cox and co2 in the lower thermosphere}.{\BBCQ}
\newblock
\APACjournalVolNumPages{Journal of Geophysical Research: Atmospheres}{121}{7}{3634--3644}.
\newblock
\begin{APACrefDOI} \doi{https://doi.org/10.1002/2015JD024553} \end{APACrefDOI}
\PrintBackRefs{\CurrentBib}

\bibitem [\protect \citeauthoryear {%
Garcia%
\ \protect \BOthers {.}}{%
Garcia%
\ \protect \BOthers {.}}{%
{\protect \APACyear {2014}}%
}]{%
garcia2014distribution}
\APACinsertmetastar {%
garcia2014distribution}%
\begin{APACrefauthors}%
Garcia, R\BPBI R.%
, L{\'o}pez-Puertas, M.%
, Funke, B.%
, Marsh, D\BPBI R.%
, Kinnison, D\BPBI E.%
, Smith, A\BPBI K.%
\BCBL {}\ \BBA {} Gonz{\'a}lez-Galindo, F.%
\end{APACrefauthors}%
\unskip\
\newblock
\APACrefYearMonthDay{2014}{}{}.
\newblock
{\BBOQ}\APACrefatitle {On the distribution of CO2 and CO in the mesosphere and lower thermosphere} {On the distribution of co2 and co in the mesosphere and lower thermosphere}.{\BBCQ}
\newblock
\APACjournalVolNumPages{Journal of Geophysical Research: Atmospheres}{119}{9}{5700--5718}.
\newblock
\begin{APACrefDOI} \doi{https://doi.org/10.1002/2013JD021208} \end{APACrefDOI}
\PrintBackRefs{\CurrentBib}

\bibitem [\protect \citeauthoryear {%
Gasperini%
}{%
Gasperini%
}{%
{\protect \APACyear {2019}}%
}]{%
WACCM_X2023}
\APACinsertmetastar {%
WACCM_X2023}%
\begin{APACrefauthors}%
Gasperini, F.%
\end{APACrefauthors}%
\unskip\
\newblock
\APACrefYearMonthDay{2019}{}{}.
\newblock
\APACrefbtitle {CCSM run SD-WACCM-X Version 1 Daily Averaged Atmosphere History Data.} {Ccsm run sd-waccm-x version 1 daily averaged atmosphere history data.}
\newblock
\APACaddressPublisher{}{Earthsystemgrid [Dataset]}.
\newblock
\begin{APACrefURL} \url{https://rda.ucar.edu/datasets/d651034/} \end{APACrefURL}
\newblock
\begin{APACrefDOI} \doi{https://doi.org/10.26024/5b58-nc53} \end{APACrefDOI}
\PrintBackRefs{\CurrentBib}

\bibitem [\protect \citeauthoryear {%
Gelaro%
\ \protect \BOthers {.}}{%
Gelaro%
\ \protect \BOthers {.}}{%
{\protect \APACyear {2017}}%
}]{%
gelaro2017modern}
\APACinsertmetastar {%
gelaro2017modern}%
\begin{APACrefauthors}%
Gelaro, R.%
, McCarty, W.%
, Su{\'a}rez, M\BPBI J.%
, Todling, R.%
, Molod, A.%
, Takacs, L.%
\BDBL {}others%
\end{APACrefauthors}%
\unskip\
\newblock
\APACrefYearMonthDay{2017}{}{}.
\newblock
{\BBOQ}\APACrefatitle {The modern-era retrospective analysis for research and applications, version 2 (MERRA-2)} {The modern-era retrospective analysis for research and applications, version 2 (merra-2)}.{\BBCQ}
\newblock
\APACjournalVolNumPages{Journal of climate}{30}{14}{5419--5454}.
\newblock
\begin{APACrefDOI} \doi{https://doi.org/10.1175/JCLI-D-16-0758.1} \end{APACrefDOI}
\PrintBackRefs{\CurrentBib}

\bibitem [\protect \citeauthoryear {%
GMAO%
}{%
GMAO%
}{%
{\protect \APACyear {2015}}%
}]{%
MERRA2_2015}
\APACinsertmetastar {%
MERRA2_2015}%
\begin{APACrefauthors}%
GMAO.%
\end{APACrefauthors}%
\unskip\
\newblock
\APACrefYearMonthDay{2015}{}{}.
\newblock
\APACrefbtitle {MERRA-2 $inst6_3d_ana_Np$: 3d,6-Hourly,Instantaneous,Pressure-Level,Analysis,Analyzed Meteorological Fields V5.12.4, Greenbelt, MD, USA, Goddard Earth Sciences Data and Information Services Center (GES DISC).} {Merra-2 $inst6_3d_ana_np$: 3d,6-hourly,instantaneous,pressure-level,analysis,analyzed meteorological fields v5.12.4, greenbelt, md, usa, goddard earth sciences data and information services center (ges disc).}
\newblock
\APACaddressPublisher{}{Earthsystemgrid [Dataset]}.
\newblock
\begin{APACrefURL} \url{https://disc.gsfc.nasa.gov/datasets/M2I6NPANA_5.12.4/summary} \end{APACrefURL}
\newblock
\begin{APACrefDOI} \doi{https://doi.org/10.5067/A7S6XP56VZWS} \end{APACrefDOI}
\PrintBackRefs{\CurrentBib}

\bibitem [\protect \citeauthoryear {%
Houghton%
}{%
Houghton%
}{%
{\protect \APACyear {1969}}%
}]{%
houghton1969absorption}
\APACinsertmetastar {%
houghton1969absorption}%
\begin{APACrefauthors}%
Houghton, J.%
\end{APACrefauthors}%
\unskip\
\newblock
\APACrefYearMonthDay{1969}{}{}.
\newblock
{\BBOQ}\APACrefatitle {Absorption and emission by carbon-dioxide in the mesosphere} {Absorption and emission by carbon-dioxide in the mesosphere}.{\BBCQ}
\newblock
\APACjournalVolNumPages{Quarterly Journal of the Royal Meteorological Society}{95}{403}{1--20}.
\newblock
\begin{APACrefDOI} \doi{https://doi.org/10.1002/qj.49709540302} \end{APACrefDOI}
\PrintBackRefs{\CurrentBib}

\bibitem [\protect \citeauthoryear {%
Khomich%
, Semenov%
\BCBL {}\ \BBA {} Shefov%
}{%
Khomich%
\ \protect \BOthers {.}}{%
{\protect \APACyear {2008}}%
}]{%
khomich2008airglow}
\APACinsertmetastar {%
khomich2008airglow}%
\begin{APACrefauthors}%
Khomich, V\BPBI Y.%
, Semenov, A\BPBI I.%
\BCBL {}\ \BBA {} Shefov, N\BPBI N.%
\end{APACrefauthors}%
\unskip\
\newblock
\APACrefYear{2008}.
\newblock
\APACrefbtitle {Airglow as an indicator of upper atmospheric structure and dynamics} {Airglow as an indicator of upper atmospheric structure and dynamics}.
\newblock
\APACaddressPublisher{}{Springer Science \& Business Media}.
\newblock
\begin{APACrefDOI} \doi{https://doi.org/10.1007/978-3-540-75833-4} \end{APACrefDOI}
\PrintBackRefs{\CurrentBib}

\bibitem [\protect \citeauthoryear {%
Kodera%
, Mukougawa%
, Maury%
, Ueda%
\BCBL {}\ \BBA {} Claud%
}{%
Kodera%
\ \protect \BOthers {.}}{%
{\protect \APACyear {2016}}%
}]{%
kodera2016absorbing}
\APACinsertmetastar {%
kodera2016absorbing}%
\begin{APACrefauthors}%
Kodera, K.%
, Mukougawa, H.%
, Maury, P.%
, Ueda, M.%
\BCBL {}\ \BBA {} Claud, C.%
\end{APACrefauthors}%
\unskip\
\newblock
\APACrefYearMonthDay{2016}{}{}.
\newblock
{\BBOQ}\APACrefatitle {Absorbing and reflecting sudden stratospheric warming events and their relationship with tropospheric circulation} {Absorbing and reflecting sudden stratospheric warming events and their relationship with tropospheric circulation}.{\BBCQ}
\newblock
\APACjournalVolNumPages{Journal of Geophysical Research: Atmospheres}{121}{1}{80--94}.
\newblock
\begin{APACrefDOI} \doi{https://doi.org/10.1002/2015JD023359} \end{APACrefDOI}
\PrintBackRefs{\CurrentBib}

\bibitem [\protect \citeauthoryear {%
Kuhn%
\ \BBA {} London%
}{%
Kuhn%
\ \BBA {} London%
}{%
{\protect \APACyear {1969}}%
}]{%
kuhn1969infrared}
\APACinsertmetastar {%
kuhn1969infrared}%
\begin{APACrefauthors}%
Kuhn, W\BPBI R.%
\BCBT {}\ \BBA {} London, J.%
\end{APACrefauthors}%
\unskip\
\newblock
\APACrefYearMonthDay{1969}{}{}.
\newblock
{\BBOQ}\APACrefatitle {Infrared radiative cooling in the middle atmosphere (30--110 km)} {Infrared radiative cooling in the middle atmosphere (30--110 km)}.{\BBCQ}
\newblock
\APACjournalVolNumPages{Journal of the Atmospheric Sciences}{26}{2}{189--204}.
\newblock
\begin{APACrefDOI} \doi{https://doi.org/10.1175/1520-0469(1969)026%3C0189:IRCITM%3E2.0.CO;2} \end{APACrefDOI}
\PrintBackRefs{\CurrentBib}

\bibitem [\protect \citeauthoryear {%
Kumar%
\ \protect \BOthers {.}}{%
Kumar%
\ \protect \BOthers {.}}{%
{\protect \APACyear {2024a}}%
}]{%
kumar2024influence}
\APACinsertmetastar {%
kumar2024influence}%
\begin{APACrefauthors}%
Kumar, A.%
, Krishna, M\BPBI V\BPBI S.%
, Ranjan, A\BPBI K.%
, Bender, S.%
, Sinnhuber, M.%
\BCBL {}\ \BBA {} Sarkhel, S.%
\end{APACrefauthors}%
\unskip\
\newblock
\APACrefYearMonthDay{2024a}{}{}.
\newblock
{\BBOQ}\APACrefatitle {Influence of temperature changes and vertically transported trace species on the structure of MLT region during major SSW events} {Influence of temperature changes and vertically transported trace species on the structure of mlt region during major ssw events}.{\BBCQ}
\newblock
\APACjournalVolNumPages{Journal of Atmospheric and Solar-Terrestrial Physics}{259}{}{106243}.
\newblock
\begin{APACrefDOI} \doi{https://doi.org/10.1016/j.jastp.2024.106243} \end{APACrefDOI}
\PrintBackRefs{\CurrentBib}

\bibitem [\protect \citeauthoryear {%
Kumar%
, Sunil~Krishna%
\BCBL {}\ \BBA {} Ranjan%
}{%
Kumar%
\ \protect \BOthers {.}}{%
{\protect \APACyear {2024b}}%
}]{%
kumar2024effect}
\APACinsertmetastar {%
kumar2024effect}%
\begin{APACrefauthors}%
Kumar, A.%
, Sunil~Krishna, M\BPBI V.%
\BCBL {}\ \BBA {} Ranjan, A\BPBI K.%
\end{APACrefauthors}%
\unskip\
\newblock
\APACrefYearMonthDay{2024b}{}{}.
\newblock
{\BBOQ}\APACrefatitle {Effect of 2009 Major SSW Event on the Mesospheric $CO_2$ Cooling} {Effect of 2009 major ssw event on the mesospheric $co_2$ cooling}.{\BBCQ}
\newblock
\APACjournalVolNumPages{Journal of Geophysical Research: Atmospheres}{129}{24}{e2024JD041298}.
\newblock
\begin{APACrefDOI} \doi{https://doi.org/10.1029/2024JD041298} \end{APACrefDOI}
\PrintBackRefs{\CurrentBib}

\bibitem [\protect \citeauthoryear {%
Kurihara%
\ \protect \BOthers {.}}{%
Kurihara%
\ \protect \BOthers {.}}{%
{\protect \APACyear {2010}}%
}]{%
kurihara2010links}
\APACinsertmetastar {%
kurihara2010links}%
\begin{APACrefauthors}%
Kurihara, J.%
, Ogawa, Y.%
, Oyama, S.%
, Nozawa, S.%
, Tsutsumi, M.%
, Hall, C.%
\BDBL {}Fujii, R.%
\end{APACrefauthors}%
\unskip\
\newblock
\APACrefYearMonthDay{2010}{}{}.
\newblock
{\BBOQ}\APACrefatitle {Links between a stratospheric sudden warming and thermal structures and dynamics in the high-latitude mesosphere, lower thermosphere, and ionosphere} {Links between a stratospheric sudden warming and thermal structures and dynamics in the high-latitude mesosphere, lower thermosphere, and ionosphere}.{\BBCQ}
\newblock
\APACjournalVolNumPages{Geophysical research letters}{37}{13}{}.
\newblock
\begin{APACrefDOI} \doi{https://doi.org/10.1029/2010GL043643} \end{APACrefDOI}
\PrintBackRefs{\CurrentBib}

\bibitem [\protect \citeauthoryear {%
Lashof%
\ \BBA {} Ahuja%
}{%
Lashof%
\ \BBA {} Ahuja%
}{%
{\protect \APACyear {1990}}%
}]{%
lashof1990relative}
\APACinsertmetastar {%
lashof1990relative}%
\begin{APACrefauthors}%
Lashof, D\BPBI A.%
\BCBT {}\ \BBA {} Ahuja, D\BPBI R.%
\end{APACrefauthors}%
\unskip\
\newblock
\APACrefYearMonthDay{1990}{}{}.
\newblock
{\BBOQ}\APACrefatitle {Relative contributions of greenhouse gas emissions to global warming} {Relative contributions of greenhouse gas emissions to global warming}.{\BBCQ}
\newblock
\APACjournalVolNumPages{Nature}{344}{6266}{529--531}.
\newblock
\begin{APACrefDOI} \doi{https://doi.org/10.1038/344529a0} \end{APACrefDOI}
\PrintBackRefs{\CurrentBib}

\bibitem [\protect \citeauthoryear {%
Laskar%
\ \protect \BOthers {.}}{%
Laskar%
\ \protect \BOthers {.}}{%
{\protect \APACyear {2019}}%
}]{%
laskar2019interhemispheric}
\APACinsertmetastar {%
laskar2019interhemispheric}%
\begin{APACrefauthors}%
Laskar, F\BPBI I.%
, McCormack, J\BPBI P.%
, Chau, J\BPBI L.%
, Pallamraju, D.%
, Hoffmann, P.%
\BCBL {}\ \BBA {} Singh, R\BPBI P.%
\end{APACrefauthors}%
\unskip\
\newblock
\APACrefYearMonthDay{2019}{}{}.
\newblock
{\BBOQ}\APACrefatitle {Interhemispheric meridional circulation during sudden stratospheric warming} {Interhemispheric meridional circulation during sudden stratospheric warming}.{\BBCQ}
\newblock
\APACjournalVolNumPages{Journal of Geophysical Research: Space Physics}{124}{8}{7112--7122}.
\newblock
\begin{APACrefDOI} \doi{https://doi.org/10.1029/2018JA026424} \end{APACrefDOI}
\PrintBackRefs{\CurrentBib}

\bibitem [\protect \citeauthoryear {%
La{\v{s}}tovi{\v{c}}ka%
}{%
La{\v{s}}tovi{\v{c}}ka%
}{%
{\protect \APACyear {2023}}%
}]{%
lavstovivcka2023progress}
\APACinsertmetastar {%
lavstovivcka2023progress}%
\begin{APACrefauthors}%
La{\v{s}}tovi{\v{c}}ka, J.%
\end{APACrefauthors}%
\unskip\
\newblock
\APACrefYearMonthDay{2023}{}{}.
\newblock
{\BBOQ}\APACrefatitle {Progress in investigating long-term trends in the mesosphere, thermosphere, and ionosphere} {Progress in investigating long-term trends in the mesosphere, thermosphere, and ionosphere}.{\BBCQ}
\newblock
\APACjournalVolNumPages{Atmospheric Chemistry and Physics}{23}{10}{5783--5800}.
\newblock
\begin{APACrefDOI} \doi{https://doi.org/10.5194/acp-23-5783-2023} \end{APACrefDOI}
\PrintBackRefs{\CurrentBib}

\bibitem [\protect \citeauthoryear {%
Limpasuvan%
, Orsolini%
, Chandran%
, Garcia%
\BCBL {}\ \BBA {} Smith%
}{%
Limpasuvan%
\ \protect \BOthers {.}}{%
{\protect \APACyear {2016}}%
}]{%
limpasuvan2016composite}
\APACinsertmetastar {%
limpasuvan2016composite}%
\begin{APACrefauthors}%
Limpasuvan, V.%
, Orsolini, Y\BPBI J.%
, Chandran, A.%
, Garcia, R\BPBI R.%
\BCBL {}\ \BBA {} Smith, A\BPBI K.%
\end{APACrefauthors}%
\unskip\
\newblock
\APACrefYearMonthDay{2016}{}{}.
\newblock
{\BBOQ}\APACrefatitle {On the composite response of the MLT to major sudden stratospheric warming events with elevated stratopause} {On the composite response of the mlt to major sudden stratospheric warming events with elevated stratopause}.{\BBCQ}
\newblock
\APACjournalVolNumPages{Journal of Geophysical Research: Atmospheres}{121}{9}{4518--4537}.
\newblock
\begin{APACrefDOI} \doi{https://doi.org/10.1002/2015JD024401} \end{APACrefDOI}
\PrintBackRefs{\CurrentBib}

\bibitem [\protect \citeauthoryear {%
Liu%
\ \protect \BOthers {.}}{%
Liu%
\ \protect \BOthers {.}}{%
{\protect \APACyear {2018}}%
}]{%
liu2018development}
\APACinsertmetastar {%
liu2018development}%
\begin{APACrefauthors}%
Liu, H\BHBI L.%
, Bardeen, C\BPBI G.%
, Foster, B\BPBI T.%
, Lauritzen, P.%
, Liu, J.%
, Lu, G.%
\BDBL {}others%
\end{APACrefauthors}%
\unskip\
\newblock
\APACrefYearMonthDay{2018}{}{}.
\newblock
{\BBOQ}\APACrefatitle {Development and validation of the Whole Atmosphere Community Climate Model with thermosphere and ionosphere extension (WACCM-X 2.0)} {Development and validation of the whole atmosphere community climate model with thermosphere and ionosphere extension (waccm-x 2.0)}.{\BBCQ}
\newblock
\APACjournalVolNumPages{Journal of Advances in Modeling Earth Systems}{10}{2}{381--402}.
\newblock
\begin{APACrefDOI} \doi{https://doi.org/10.1002/2017MS001232} \end{APACrefDOI}
\PrintBackRefs{\CurrentBib}

\bibitem [\protect \citeauthoryear {%
Liu%
\ \BBA {} Roble%
}{%
Liu%
\ \BBA {} Roble%
}{%
{\protect \APACyear {2002}}%
}]{%
liu2002study}
\APACinsertmetastar {%
liu2002study}%
\begin{APACrefauthors}%
Liu, H\BHBI L.%
\BCBT {}\ \BBA {} Roble, R.%
\end{APACrefauthors}%
\unskip\
\newblock
\APACrefYearMonthDay{2002}{}{}.
\newblock
{\BBOQ}\APACrefatitle {A study of a self-generated stratospheric sudden warming and its mesospheric--lower thermospheric impacts using the coupled TIME-GCM/CCM3} {A study of a self-generated stratospheric sudden warming and its mesospheric--lower thermospheric impacts using the coupled time-gcm/ccm3}.{\BBCQ}
\newblock
\APACjournalVolNumPages{Journal of Geophysical Research: Atmospheres}{107}{D23}{ACL--15}.
\newblock
\begin{APACrefDOI} \doi{https://doi.org/10.1029/2001JD001533} \end{APACrefDOI}
\PrintBackRefs{\CurrentBib}

\bibitem [\protect \citeauthoryear {%
L{\'o}pez-Puertas%
, L{\'o}pez-Valverde%
, Garcia%
\BCBL {}\ \BBA {} Roble%
}{%
L{\'o}pez-Puertas%
\ \protect \BOthers {.}}{%
{\protect \APACyear {2000}}%
}]{%
lopez2000review}
\APACinsertmetastar {%
lopez2000review}%
\begin{APACrefauthors}%
L{\'o}pez-Puertas, M.%
, L{\'o}pez-Valverde, M\BPBI {\'A}.%
, Garcia, R\BPBI R.%
\BCBL {}\ \BBA {} Roble, R\BPBI G.%
\end{APACrefauthors}%
\unskip\
\newblock
\APACrefYearMonthDay{2000}{}{}.
\newblock
{\BBOQ}\APACrefatitle {A review of CO 2 and CO abundances in the middle atmosphere} {A review of co 2 and co abundances in the middle atmosphere}.{\BBCQ}
\newblock
\APACjournalVolNumPages{Washington DC American Geophysical Union Geophysical Monograph Series}{123}{}{83--100}.
\newblock
\begin{APACrefDOI} \doi{https://doi.org/10.1029/GM123p0083} \end{APACrefDOI}
\PrintBackRefs{\CurrentBib}

\bibitem [\protect \citeauthoryear {%
L{\'o}pez-Puertas%
\ \BBA {} Taylor%
}{%
L{\'o}pez-Puertas%
\ \BBA {} Taylor%
}{%
{\protect \APACyear {2001}}%
}]{%
lopez2001non}
\APACinsertmetastar {%
lopez2001non}%
\begin{APACrefauthors}%
L{\'o}pez-Puertas, M.%
\BCBT {}\ \BBA {} Taylor, F\BPBI W.%
\end{APACrefauthors}%
\unskip\
\newblock
\APACrefYear{2001}.
\newblock
\APACrefbtitle {Non-LTE radiative transfer in the Atmosphere} {Non-lte radiative transfer in the atmosphere}\ (\BVOL~3).
\newblock
\APACaddressPublisher{}{World Scientific}.
\newblock
\begin{APACrefDOI} \doi{https://doi.org/10.1142/4650} \end{APACrefDOI}
\PrintBackRefs{\CurrentBib}

\bibitem [\protect \citeauthoryear {%
Manney%
\ \protect \BOthers {.}}{%
Manney%
\ \protect \BOthers {.}}{%
{\protect \APACyear {2009}}%
}]{%
manney2009aura}
\APACinsertmetastar {%
manney2009aura}%
\begin{APACrefauthors}%
Manney, G\BPBI L.%
, Schwartz, M\BPBI J.%
, Kr{\"u}ger, K.%
, Santee, M\BPBI L.%
, Pawson, S.%
, Lee, J\BPBI N.%
\BDBL {}Livesey, N\BPBI J.%
\end{APACrefauthors}%
\unskip\
\newblock
\APACrefYearMonthDay{2009}{}{}.
\newblock
{\BBOQ}\APACrefatitle {Aura Microwave Limb Sounder observations of dynamics and transport during the record-breaking 2009 Arctic stratospheric major warming} {Aura microwave limb sounder observations of dynamics and transport during the record-breaking 2009 arctic stratospheric major warming}.{\BBCQ}
\newblock
\APACjournalVolNumPages{Geophysical Research Letters}{36}{12}{}.
\newblock
\begin{APACrefDOI} \doi{https://doi.org/10.1029/2009GL038586} \end{APACrefDOI}
\PrintBackRefs{\CurrentBib}

\bibitem [\protect \citeauthoryear {%
Matsuno%
}{%
Matsuno%
}{%
{\protect \APACyear {1971}}%
}]{%
matsuno1971dynamical}
\APACinsertmetastar {%
matsuno1971dynamical}%
\begin{APACrefauthors}%
Matsuno, T.%
\end{APACrefauthors}%
\unskip\
\newblock
\APACrefYearMonthDay{1971}{}{}.
\newblock
{\BBOQ}\APACrefatitle {A dynamical model of the stratospheric sudden warming} {A dynamical model of the stratospheric sudden warming}.{\BBCQ}
\newblock
\APACjournalVolNumPages{Journal of Atmospheric Sciences}{28}{8}{1479--1494}.
\newblock
\begin{APACrefDOI} \doi{https://doi.org/10.1175/1520-0469(1971)028<1479:ADMOTS>2.0.CO;2} \end{APACrefDOI}
\PrintBackRefs{\CurrentBib}

\bibitem [\protect \citeauthoryear {%
Mlynczak%
}{%
Mlynczak%
}{%
{\protect \APACyear {2000}}%
}]{%
mlynczak2000contemporary}
\APACinsertmetastar {%
mlynczak2000contemporary}%
\begin{APACrefauthors}%
Mlynczak, M\BPBI G.%
\end{APACrefauthors}%
\unskip\
\newblock
\APACrefYearMonthDay{2000}{}{}.
\newblock
{\BBOQ}\APACrefatitle {A contemporary assessment of the mesospheric energy budget} {A contemporary assessment of the mesospheric energy budget}.{\BBCQ}
\newblock
\APACjournalVolNumPages{Washington DC American Geophysical Union Geophysical Monograph Series}{123}{}{37--52}.
\newblock
\begin{APACrefDOI} \doi{https://doi.org/10.1029/GM123p0037} \end{APACrefDOI}
\PrintBackRefs{\CurrentBib}

\bibitem [\protect \citeauthoryear {%
Mlynczak%
\ \protect \BOthers {.}}{%
Mlynczak%
\ \protect \BOthers {.}}{%
{\protect \APACyear {2022}}%
}]{%
mlynczak2022cooling}
\APACinsertmetastar {%
mlynczak2022cooling}%
\begin{APACrefauthors}%
Mlynczak, M\BPBI G.%
, Hunt, L\BPBI A.%
, Garcia, R\BPBI R.%
, Harvey, V\BPBI L.%
, Marshall, B\BPBI T.%
, Yue, J.%
\BDBL {}Russell~III, J\BPBI M.%
\end{APACrefauthors}%
\unskip\
\newblock
\APACrefYearMonthDay{2022}{}{}.
\newblock
{\BBOQ}\APACrefatitle {Cooling and Contraction of the Mesosphere and Lower Thermosphere from 2002 to 2021} {Cooling and contraction of the mesosphere and lower thermosphere from 2002 to 2021}.{\BBCQ}
\newblock
\APACjournalVolNumPages{Journal of Geophysical Research: Atmospheres}{127}{22}{e2022JD036767}.
\newblock
\begin{APACrefDOI} \doi{https://doi.org/10.1029/2022JD036767} \end{APACrefDOI}
\PrintBackRefs{\CurrentBib}

\bibitem [\protect \citeauthoryear {%
Mlynczak%
\ \protect \BOthers {.}}{%
Mlynczak%
\ \protect \BOthers {.}}{%
{\protect \APACyear {2010}}%
}]{%
mlynczak2010observations}
\APACinsertmetastar {%
mlynczak2010observations}%
\begin{APACrefauthors}%
Mlynczak, M\BPBI G.%
, Hunt, L\BPBI A.%
, Thomas~Marshall, B.%
, Martin-Torres, F\BPBI J.%
, Mertens, C\BPBI J.%
, Russell~III, J\BPBI M.%
\BDBL {}others%
\end{APACrefauthors}%
\unskip\
\newblock
\APACrefYearMonthDay{2010}{}{}.
\newblock
{\BBOQ}\APACrefatitle {Observations of infrared radiative cooling in the thermosphere on daily to multiyear timescales from the TIMED/SABER instrument} {Observations of infrared radiative cooling in the thermosphere on daily to multiyear timescales from the timed/saber instrument}.{\BBCQ}
\newblock
\APACjournalVolNumPages{Journal of Geophysical Research: Space Physics}{115}{A3}{}.
\newblock
\begin{APACrefDOI} \doi{https://doi.org/10.1029/2009JA014713} \end{APACrefDOI}
\PrintBackRefs{\CurrentBib}

\bibitem [\protect \citeauthoryear {%
Nath%
, Chen%
, Zelin%
, Pogoreltsev%
\BCBL {}\ \BBA {} Wei%
}{%
Nath%
\ \protect \BOthers {.}}{%
{\protect \APACyear {2016}}%
}]{%
nath2016dynamics}
\APACinsertmetastar {%
nath2016dynamics}%
\begin{APACrefauthors}%
Nath, D.%
, Chen, W.%
, Zelin, C.%
, Pogoreltsev, A\BPBI I.%
\BCBL {}\ \BBA {} Wei, K.%
\end{APACrefauthors}%
\unskip\
\newblock
\APACrefYearMonthDay{2016}{}{}.
\newblock
{\BBOQ}\APACrefatitle {Dynamics of 2013 Sudden Stratospheric Warming event and its impact on cold weather over Eurasia: Role of planetary wave reflection} {Dynamics of 2013 sudden stratospheric warming event and its impact on cold weather over eurasia: Role of planetary wave reflection}.{\BBCQ}
\newblock
\APACjournalVolNumPages{Scientific reports}{6}{1}{1--12}.
\newblock
\begin{APACrefDOI} \doi{https://doi.org/10.1038/srep24174} \end{APACrefDOI}
\PrintBackRefs{\CurrentBib}

\bibitem [\protect \citeauthoryear {%
Nayak%
\ \BBA {} Yi{\u{g}}it%
}{%
Nayak%
\ \BBA {} Yi{\u{g}}it%
}{%
{\protect \APACyear {2019}}%
}]{%
nayak2019variation}
\APACinsertmetastar {%
nayak2019variation}%
\begin{APACrefauthors}%
Nayak, C.%
\BCBT {}\ \BBA {} Yi{\u{g}}it, E.%
\end{APACrefauthors}%
\unskip\
\newblock
\APACrefYearMonthDay{2019}{}{}.
\newblock
{\BBOQ}\APACrefatitle {Variation of small-scale gravity wave activity in the ionosphere during the major sudden stratospheric warming event of 2009} {Variation of small-scale gravity wave activity in the ionosphere during the major sudden stratospheric warming event of 2009}.{\BBCQ}
\newblock
\APACjournalVolNumPages{Journal of Geophysical Research: Space Physics}{124}{1}{470--488}.
\newblock
\begin{APACrefDOI} \doi{https://doi.org/10.1029/2018JA026048} \end{APACrefDOI}
\PrintBackRefs{\CurrentBib}

\bibitem [\protect \citeauthoryear {%
Orsolini%
, Zhang%
\BCBL {}\ \BBA {} Limpasuvan%
}{%
Orsolini%
\ \protect \BOthers {.}}{%
{\protect \APACyear {2022}}%
}]{%
orsolini2022abrupt}
\APACinsertmetastar {%
orsolini2022abrupt}%
\begin{APACrefauthors}%
Orsolini, Y\BPBI J.%
, Zhang, J.%
\BCBL {}\ \BBA {} Limpasuvan, V.%
\end{APACrefauthors}%
\unskip\
\newblock
\APACrefYearMonthDay{2022}{}{}.
\newblock
{\BBOQ}\APACrefatitle {Abrupt change in the lower thermospheric mean meridional circulation during sudden stratospheric warmings and its impact on trace species} {Abrupt change in the lower thermospheric mean meridional circulation during sudden stratospheric warmings and its impact on trace species}.{\BBCQ}
\newblock
\APACjournalVolNumPages{Journal of Geophysical Research: Atmospheres}{127}{20}{e2022JD037050}.
\newblock
\begin{APACrefDOI} \doi{https://doi.org/10.1029/2022JD037050} \end{APACrefDOI}
\PrintBackRefs{\CurrentBib}

\bibitem [\protect \citeauthoryear {%
Qian%
, Burns%
, Solomon%
\BCBL {}\ \BBA {} Wang%
}{%
Qian%
\ \protect \BOthers {.}}{%
{\protect \APACyear {2017}}%
}]{%
qian2017carbon}
\APACinsertmetastar {%
qian2017carbon}%
\begin{APACrefauthors}%
Qian, L.%
, Burns, A\BPBI G.%
, Solomon, S\BPBI C.%
\BCBL {}\ \BBA {} Wang, W.%
\end{APACrefauthors}%
\unskip\
\newblock
\APACrefYearMonthDay{2017}{}{}.
\newblock
{\BBOQ}\APACrefatitle {Carbon dioxide trends in the mesosphere and lower thermosphere} {Carbon dioxide trends in the mesosphere and lower thermosphere}.{\BBCQ}
\newblock
\APACjournalVolNumPages{Journal of Geophysical Research: Space Physics}{122}{4}{4474--4488}.
\newblock
\begin{APACrefDOI} \doi{https://doi.org/10.1002/2016JA023825} \end{APACrefDOI}
\PrintBackRefs{\CurrentBib}

\bibitem [\protect \citeauthoryear {%
Randall%
\ \protect \BOthers {.}}{%
Randall%
\ \protect \BOthers {.}}{%
{\protect \APACyear {2006}}%
}]{%
randall2006enhanced}
\APACinsertmetastar {%
randall2006enhanced}%
\begin{APACrefauthors}%
Randall, C.%
, Harvey, V.%
, Singleton, C.%
, Bernath, P.%
, Boone, C.%
\BCBL {}\ \BBA {} Kozyra, J.%
\end{APACrefauthors}%
\unskip\
\newblock
\APACrefYearMonthDay{2006}{}{}.
\newblock
{\BBOQ}\APACrefatitle {Enhanced NOx in 2006 linked to strong upper stratospheric Arctic vortex} {Enhanced nox in 2006 linked to strong upper stratospheric arctic vortex}.{\BBCQ}
\newblock
\APACjournalVolNumPages{Geophysical Research Letters}{33}{18}{}.
\newblock
\begin{APACrefDOI} \doi{https://doi.org/10.1029/2006GL027160} \end{APACrefDOI}
\PrintBackRefs{\CurrentBib}

\bibitem [\protect \citeauthoryear {%
Rienecker%
\ \protect \BOthers {.}}{%
Rienecker%
\ \protect \BOthers {.}}{%
{\protect \APACyear {2011}}%
}]{%
rienecker2011merra}
\APACinsertmetastar {%
rienecker2011merra}%
\begin{APACrefauthors}%
Rienecker, M\BPBI M.%
, Suarez, M\BPBI J.%
, Gelaro, R.%
, Todling, R.%
, Bacmeister, J.%
, Liu, E.%
\BDBL {}others%
\end{APACrefauthors}%
\unskip\
\newblock
\APACrefYearMonthDay{2011}{}{}.
\newblock
{\BBOQ}\APACrefatitle {MERRA: NASA’s modern-era retrospective analysis for research and applications} {Merra: Nasa’s modern-era retrospective analysis for research and applications}.{\BBCQ}
\newblock
\APACjournalVolNumPages{Journal of climate}{24}{14}{3624--3648}.
\newblock
\begin{APACrefDOI} \doi{https://doi.org/10.1175/JCLI-D-11-00015.1} \end{APACrefDOI}
\PrintBackRefs{\CurrentBib}

\bibitem [\protect \citeauthoryear {%
Roble%
}{%
Roble%
}{%
{\protect \APACyear {1993}}%
}]{%
roble1993greenhouse}
\APACinsertmetastar {%
roble1993greenhouse}%
\begin{APACrefauthors}%
Roble, R.%
\end{APACrefauthors}%
\unskip\
\newblock
\APACrefYearMonthDay{1993}{}{}.
\newblock
\APACrefbtitle {“Greenhouse cooling” of the upper atmosphere.} {“greenhouse cooling” of the upper atmosphere.}
\newblock
\APACaddressPublisher{}{Wiley Online Library}.
\newblock
\begin{APACrefDOI} \doi{https://doi.org/10.1029/93EO00233} \end{APACrefDOI}
\PrintBackRefs{\CurrentBib}

\bibitem [\protect \citeauthoryear {%
Roble%
\ \BBA {} Dickinson%
}{%
Roble%
\ \BBA {} Dickinson%
}{%
{\protect \APACyear {1989}}%
}]{%
roble1989will}
\APACinsertmetastar {%
roble1989will}%
\begin{APACrefauthors}%
Roble, R.%
\BCBT {}\ \BBA {} Dickinson, R.%
\end{APACrefauthors}%
\unskip\
\newblock
\APACrefYearMonthDay{1989}{}{}.
\newblock
{\BBOQ}\APACrefatitle {How will changes in carbon dioxide and methane modify the mean structure of the mesosphere and thermosphere?} {How will changes in carbon dioxide and methane modify the mean structure of the mesosphere and thermosphere?}{\BBCQ}
\newblock
\APACjournalVolNumPages{Geophysical Research Letters}{16}{12}{1441--1444}.
\newblock
\begin{APACrefDOI} \doi{https://doi.org/10.1029/GL016i012p01441} \end{APACrefDOI}
\PrintBackRefs{\CurrentBib}

\bibitem [\protect \citeauthoryear {%
Russell~III%
, Mlynczak%
, Gordley%
, Tansock~Jr%
\BCBL {}\ \BBA {} Esplin%
}{%
Russell~III%
\ \protect \BOthers {.}}{%
{\protect \APACyear {1999}}%
}]{%
russell1999overview}
\APACinsertmetastar {%
russell1999overview}%
\begin{APACrefauthors}%
Russell~III, J\BPBI M.%
, Mlynczak, M\BPBI G.%
, Gordley, L\BPBI L.%
, Tansock~Jr, J\BPBI J.%
\BCBL {}\ \BBA {} Esplin, R\BPBI W.%
\end{APACrefauthors}%
\unskip\
\newblock
\APACrefYearMonthDay{1999}{}{}.
\newblock
{\BBOQ}\APACrefatitle {Overview of the SABER experiment and preliminary calibration results} {Overview of the saber experiment and preliminary calibration results}.{\BBCQ}
\newblock
\BIn{} \APACrefbtitle {Optical spectroscopic techniques and instrumentation for atmospheric and space research III} {Optical spectroscopic techniques and instrumentation for atmospheric and space research iii}\ (\BVOL\ 3756, \BPGS\ 277--288).
\newblock
\begin{APACrefDOI} \doi{https://doi.org/10.1117/12.366382} \end{APACrefDOI}
\PrintBackRefs{\CurrentBib}

\bibitem [\protect \citeauthoryear {%
SABER-Team%
}{%
SABER-Team%
}{%
{\protect \APACyear {2002}}%
}]{%
SABER2023}
\APACinsertmetastar {%
SABER2023}%
\begin{APACrefauthors}%
SABER-Team.%
\end{APACrefauthors}%
\unskip\
\newblock
\APACrefYearMonthDay{2002}{}{}.
\newblock
\APACrefbtitle {{Sounding of the Atmosphere using Broadband Emission Radiometry (SABER) Data from TIMED}.} {{Sounding of the Atmosphere using Broadband Emission Radiometry (SABER) Data from TIMED}.}
\newblock
\APACaddressPublisher{}{SABER [Dataset]}.
\newblock
\begin{APACrefURL} \url{http://gats-inc.com/projects_saber.htm} \end{APACrefURL}
\PrintBackRefs{\CurrentBib}

\bibitem [\protect \citeauthoryear {%
Salmi%
\ \protect \BOthers {.}}{%
Salmi%
\ \protect \BOthers {.}}{%
{\protect \APACyear {2011}}%
}]{%
salmi2011mesosphere}
\APACinsertmetastar {%
salmi2011mesosphere}%
\begin{APACrefauthors}%
Salmi, S\BHBI M.%
, Verronen, P.%
, Th{\"o}lix, L.%
, Kyr{\"o}l{\"a}, E.%
, Backman, L.%
, Karpechko, A\BPBI Y.%
\BCBL {}\ \BBA {} Sepp{\"a}l{\"a}, A.%
\end{APACrefauthors}%
\unskip\
\newblock
\APACrefYearMonthDay{2011}{}{}.
\newblock
{\BBOQ}\APACrefatitle {Mesosphere-to-stratosphere descent of odd nitrogen in February--March 2009 after sudden stratospheric warming} {Mesosphere-to-stratosphere descent of odd nitrogen in february--march 2009 after sudden stratospheric warming}.{\BBCQ}
\newblock
\APACjournalVolNumPages{Atmospheric Chemistry and Physics}{11}{10}{4645--4655}.
\newblock
\begin{APACrefDOI} \doi{https://doi.org/10.5194/acp-11-4645-2011} \end{APACrefDOI}
\PrintBackRefs{\CurrentBib}

\bibitem [\protect \citeauthoryear {%
Sharma%
\ \BBA {} Wintersteiner%
}{%
Sharma%
\ \BBA {} Wintersteiner%
}{%
{\protect \APACyear {1990}}%
}]{%
sharma1990role}
\APACinsertmetastar {%
sharma1990role}%
\begin{APACrefauthors}%
Sharma, R\BPBI D.%
\BCBT {}\ \BBA {} Wintersteiner, P\BPBI P.%
\end{APACrefauthors}%
\unskip\
\newblock
\APACrefYearMonthDay{1990}{}{}.
\newblock
{\BBOQ}\APACrefatitle {Role of carbon dioxide in cooling planetary thermospheres} {Role of carbon dioxide in cooling planetary thermospheres}.{\BBCQ}
\newblock
\APACjournalVolNumPages{Geophysical Research Letters}{17}{12}{2201--2204}.
\newblock
\begin{APACrefDOI} \doi{https://doi.org/10.1029/GL017i012p02201} \end{APACrefDOI}
\PrintBackRefs{\CurrentBib}

\bibitem [\protect \citeauthoryear {%
M.~Shepherd%
, Beagley%
\BCBL {}\ \BBA {} Fomichev%
}{%
M.~Shepherd%
\ \protect \BOthers {.}}{%
{\protect \APACyear {2014}}%
}]{%
shepherd2014stratospheric}
\APACinsertmetastar {%
shepherd2014stratospheric}%
\begin{APACrefauthors}%
Shepherd, M.%
, Beagley, S.%
\BCBL {}\ \BBA {} Fomichev, V.%
\end{APACrefauthors}%
\unskip\
\newblock
\APACrefYearMonthDay{2014}{}{}.
\newblock
{\BBOQ}\APACrefatitle {Stratospheric warming influence on the mesosphere/lower thermosphere as seen by the extended CMAM} {Stratospheric warming influence on the mesosphere/lower thermosphere as seen by the extended cmam}.{\BBCQ}
\newblock
\BIn{} \APACrefbtitle {Annales Geophysicae} {Annales geophysicae}\ (\BVOL~32, \BPGS\ 589--608).
\newblock
\begin{APACrefDOI} \doi{https://doi.org/10.5194/angeo-32-589-2014} \end{APACrefDOI}
\PrintBackRefs{\CurrentBib}

\bibitem [\protect \citeauthoryear {%
M\BPBI G.~Shepherd%
, Cho%
, Shepherd%
, Ward%
\BCBL {}\ \BBA {} Drummond%
}{%
M\BPBI G.~Shepherd%
\ \protect \BOthers {.}}{%
{\protect \APACyear {2010}}%
}]{%
shepherd2010mesospheric}
\APACinsertmetastar {%
shepherd2010mesospheric}%
\begin{APACrefauthors}%
Shepherd, M\BPBI G.%
, Cho, Y\BHBI M.%
, Shepherd, G\BPBI G.%
, Ward, W.%
\BCBL {}\ \BBA {} Drummond, J\BPBI R.%
\end{APACrefauthors}%
\unskip\
\newblock
\APACrefYearMonthDay{2010}{}{}.
\newblock
{\BBOQ}\APACrefatitle {Mesospheric temperature and atomic oxygen response during the January 2009 major stratospheric warming} {Mesospheric temperature and atomic oxygen response during the january 2009 major stratospheric warming}.{\BBCQ}
\newblock
\APACjournalVolNumPages{Journal of Geophysical Research: Space Physics}{115}{A7}{}.
\newblock
\begin{APACrefDOI} \doi{https://doi.org/10.1029/2009JA015172} \end{APACrefDOI}
\PrintBackRefs{\CurrentBib}

\bibitem [\protect \citeauthoryear {%
Shved%
, Khvorostovskaya%
, Potekhin%
, Ogibalov%
\BCBL {}\ \BBA {} Uzyukova%
}{%
Shved%
\ \protect \BOthers {.}}{%
{\protect \APACyear {2003}}%
}]{%
shved2003measurement}
\APACinsertmetastar {%
shved2003measurement}%
\begin{APACrefauthors}%
Shved, G\BPBI M.%
, Khvorostovskaya, L\BPBI E.%
, Potekhin, I\BPBI Y.%
, Ogibalov, V\BPBI P.%
\BCBL {}\ \BBA {} Uzyukova, T\BPBI V.%
\end{APACrefauthors}%
\unskip\
\newblock
\APACrefYearMonthDay{2003}{}{}.
\newblock
{\BBOQ}\APACrefatitle {Measurement of rate constant for quenching CO2 (0110) by atomic oxygen at low temperatures: reassessment of the population of CO2 (0110) and the CO2 15-um emission cooling in the lower thermosphere} {Measurement of rate constant for quenching co2 (0110) by atomic oxygen at low temperatures: reassessment of the population of co2 (0110) and the co2 15-um emission cooling in the lower thermosphere}.{\BBCQ}
\newblock
\BIn{} \APACrefbtitle {Remote Sensing of Clouds and the Atmosphere VII} {Remote sensing of clouds and the atmosphere vii}\ (\BVOL\ 4882, \BPGS\ 106--116).
\newblock
\begin{APACrefDOI} \doi{https://doi.org/10.1117/12.463371} \end{APACrefDOI}
\PrintBackRefs{\CurrentBib}

\bibitem [\protect \citeauthoryear {%
Siskind%
, Coy%
\BCBL {}\ \BBA {} Espy%
}{%
Siskind%
\ \protect \BOthers {.}}{%
{\protect \APACyear {2005}}%
}]{%
siskind2005observations}
\APACinsertmetastar {%
siskind2005observations}%
\begin{APACrefauthors}%
Siskind, D.%
, Coy, L.%
\BCBL {}\ \BBA {} Espy, P.%
\end{APACrefauthors}%
\unskip\
\newblock
\APACrefYearMonthDay{2005}{}{}.
\newblock
{\BBOQ}\APACrefatitle {Observations of stratospheric warmings and mesospheric coolings by the TIMED SABER instrument} {Observations of stratospheric warmings and mesospheric coolings by the timed saber instrument}.{\BBCQ}
\newblock
\APACjournalVolNumPages{Geophysical Research Letters}{32}{9}{}.
\newblock
\begin{APACrefDOI} \doi{https://doi.org/10.1029/2005GL022399} \end{APACrefDOI}
\PrintBackRefs{\CurrentBib}

\bibitem [\protect \citeauthoryear {%
Solomon%
, Plattner%
, Knutti%
\BCBL {}\ \BBA {} Friedlingstein%
}{%
Solomon%
\ \protect \BOthers {.}}{%
{\protect \APACyear {2009}}%
}]{%
solomon2009irreversible}
\APACinsertmetastar {%
solomon2009irreversible}%
\begin{APACrefauthors}%
Solomon, S.%
, Plattner, G\BHBI K.%
, Knutti, R.%
\BCBL {}\ \BBA {} Friedlingstein, P.%
\end{APACrefauthors}%
\unskip\
\newblock
\APACrefYearMonthDay{2009}{}{}.
\newblock
{\BBOQ}\APACrefatitle {Irreversible climate change due to carbon dioxide emissions} {Irreversible climate change due to carbon dioxide emissions}.{\BBCQ}
\newblock
\APACjournalVolNumPages{Proceedings of the national academy of sciences}{106}{6}{1704--1709}.
\newblock
\begin{APACrefDOI} \doi{https://doi.org/10.1073/pnas.0812721106} \end{APACrefDOI}
\PrintBackRefs{\CurrentBib}

\bibitem [\protect \citeauthoryear {%
Soucy%
, Chateauneuf%
, Deutsch%
\BCBL {}\ \BBA {} Etienne%
}{%
Soucy%
\ \protect \BOthers {.}}{%
{\protect \APACyear {2002}}%
}]{%
soucy2002ace}
\APACinsertmetastar {%
soucy2002ace}%
\begin{APACrefauthors}%
Soucy, M\BHBI A\BPBI A.%
, Chateauneuf, F.%
, Deutsch, C.%
\BCBL {}\ \BBA {} Etienne, N.%
\end{APACrefauthors}%
\unskip\
\newblock
\APACrefYearMonthDay{2002}{}{}.
\newblock
{\BBOQ}\APACrefatitle {ACE-FTS instrument detailed design} {Ace-fts instrument detailed design}.{\BBCQ}
\newblock
\BIn{} \APACrefbtitle {Earth Observing Systems VII} {Earth observing systems vii}\ (\BVOL\ 4814, \BPGS\ 70--81).
\newblock
\begin{APACrefDOI} \doi{https://doi.org/10.1117/12.451701} \end{APACrefDOI}
\PrintBackRefs{\CurrentBib}

\bibitem [\protect \citeauthoryear {%
Tweedy%
\ \protect \BOthers {.}}{%
Tweedy%
\ \protect \BOthers {.}}{%
{\protect \APACyear {2013}}%
}]{%
tweedy2013nighttime}
\APACinsertmetastar {%
tweedy2013nighttime}%
\begin{APACrefauthors}%
Tweedy, O\BPBI V.%
, Limpasuvan, V.%
, Orsolini, Y\BPBI J.%
, Smith, A\BPBI K.%
, Garcia, R\BPBI R.%
, Kinnison, D.%
\BDBL {}others%
\end{APACrefauthors}%
\unskip\
\newblock
\APACrefYearMonthDay{2013}{}{}.
\newblock
{\BBOQ}\APACrefatitle {Nighttime secondary ozone layer during major stratospheric sudden warmings in specified-dynamics WACCM} {Nighttime secondary ozone layer during major stratospheric sudden warmings in specified-dynamics waccm}.{\BBCQ}
\newblock
\APACjournalVolNumPages{Journal of Geophysical Research: Atmospheres}{118}{15}{8346--8358}.
\newblock
\begin{APACrefDOI} \doi{https://doi.org/10.1002/jgrd.50651} \end{APACrefDOI}
\PrintBackRefs{\CurrentBib}

\bibitem [\protect \citeauthoryear {%
N.~Wang%
\ \protect \BOthers {.}}{%
N.~Wang%
\ \protect \BOthers {.}}{%
{\protect \APACyear {2022}}%
}]{%
wang2022climatology}
\APACinsertmetastar {%
wang2022climatology}%
\begin{APACrefauthors}%
Wang, N.%
, Qian, L.%
, Yue, J.%
, Wang, W.%
, Mlynczak, M\BPBI G.%
\BCBL {}\ \BBA {} Russell~III, J\BPBI M.%
\end{APACrefauthors}%
\unskip\
\newblock
\APACrefYearMonthDay{2022}{}{}.
\newblock
{\BBOQ}\APACrefatitle {Climatology of Mesosphere and Lower Thermosphere Residual Circulations and Mesopause Height Derived From SABER Observations} {Climatology of mesosphere and lower thermosphere residual circulations and mesopause height derived from saber observations}.{\BBCQ}
\newblock
\APACjournalVolNumPages{Journal of Geophysical Research: Atmospheres}{127}{4}{e2021JD035666}.
\newblock
\begin{APACrefDOI} \doi{https://doi.org/10.1029/2021JD035666} \end{APACrefDOI}
\PrintBackRefs{\CurrentBib}

\bibitem [\protect \citeauthoryear {%
Y.~Wang%
\ \protect \BOthers {.}}{%
Y.~Wang%
\ \protect \BOthers {.}}{%
{\protect \APACyear {2019}}%
}]{%
wang2019winter}
\APACinsertmetastar {%
wang2019winter}%
\begin{APACrefauthors}%
Wang, Y.%
, Shulga, V.%
, Milinevsky, G.%
, Patoka, A.%
, Evtushevsky, O.%
, Klekociuk, A.%
\BDBL {}others%
\end{APACrefauthors}%
\unskip\
\newblock
\APACrefYearMonthDay{2019}{}{}.
\newblock
{\BBOQ}\APACrefatitle {Winter 2018 major sudden stratospheric warming impact on midlatitude mesosphere from microwave radiometer measurements} {Winter 2018 major sudden stratospheric warming impact on midlatitude mesosphere from microwave radiometer measurements}.{\BBCQ}
\newblock
\APACjournalVolNumPages{Atmospheric Chemistry and Physics}{19}{15}{10303--10317}.
\newblock
\begin{APACrefDOI} \doi{https://doi.org/10.5194/acp-19-10303-2019} \end{APACrefDOI}
\PrintBackRefs{\CurrentBib}

\bibitem [\protect \citeauthoryear {%
Yi{\u{g}}it%
, Kn{\'\i}{\v{z}}ov{\'a}%
, Georgieva%
\BCBL {}\ \BBA {} Ward%
}{%
Yi{\u{g}}it%
\ \protect \BOthers {.}}{%
{\protect \APACyear {2016}}%
}]{%
yiugit2016review}
\APACinsertmetastar {%
yiugit2016review}%
\begin{APACrefauthors}%
Yi{\u{g}}it, E.%
, Kn{\'\i}{\v{z}}ov{\'a}, P\BPBI K.%
, Georgieva, K.%
\BCBL {}\ \BBA {} Ward, W.%
\end{APACrefauthors}%
\unskip\
\newblock
\APACrefYearMonthDay{2016}{}{}.
\newblock
{\BBOQ}\APACrefatitle {A review of vertical coupling in the Atmosphere--Ionosphere system: Effects of waves, sudden stratospheric warmings, space weather, and of solar activity} {A review of vertical coupling in the atmosphere--ionosphere system: Effects of waves, sudden stratospheric warmings, space weather, and of solar activity}.{\BBCQ}
\newblock
\APACjournalVolNumPages{Journal of Atmospheric and Solar-Terrestrial Physics}{141}{}{1--12}.
\newblock
\begin{APACrefDOI} \doi{https://doi.org/10.1016/j.jastp.2016.02.011} \end{APACrefDOI}
\PrintBackRefs{\CurrentBib}

\bibitem [\protect \citeauthoryear {%
Z{\"u}licke%
\ \BBA {} Becker%
}{%
Z{\"u}licke%
\ \BBA {} Becker%
}{%
{\protect \APACyear {2013}}%
}]{%
zulicke2013structure}
\APACinsertmetastar {%
zulicke2013structure}%
\begin{APACrefauthors}%
Z{\"u}licke, C.%
\BCBT {}\ \BBA {} Becker, E.%
\end{APACrefauthors}%
\unskip\
\newblock
\APACrefYearMonthDay{2013}{}{}.
\newblock
{\BBOQ}\APACrefatitle {The structure of the mesosphere during sudden stratospheric warmings in a global circulation model} {The structure of the mesosphere during sudden stratospheric warmings in a global circulation model}.{\BBCQ}
\newblock
\APACjournalVolNumPages{Journal of Geophysical Research: Atmospheres}{118}{5}{2255--2271}.
\newblock
\begin{APACrefDOI} \doi{https://doi.org/10.1002/jgrd.50219} \end{APACrefDOI}
\PrintBackRefs{\CurrentBib}

\end{thebibliography}
 
%


%
%
%
%
%

\end{document}